\documentclass[journal]{IEEEtran}
\ifCLASSINFOpdf

\else
 \fi
\usepackage{epsfig,rotating,setspace,latexsym,amsmath,epsf,amssymb,bm}
\usepackage{cite,authblk,color}

\begin{document}
\title{Degrees of Freedom Region of the MIMO Interference Channel with Output Feedback and Delayed CSIT}

\author{Ravi~Tandon,~\IEEEmembership{Member,~IEEE,}
        Soheil~Mohajer,~\IEEEmembership{Member,~IEEE,}
        H.~Vincent~Poor,~\IEEEmembership{Fellow,~IEEE,}
         and~Shlomo~Shamai,~\IEEEmembership{Fellow,~IEEE}%
\thanks{Manuscript received September 2011, revised October 2012. }%
\thanks{Ravi Tandon is with the Department of Electrical and Computer Engineering, Virginia Tech, Blacksburg, VA, USA. E-mail: tandonr@vt.edu.}%
\thanks{ Soheil Mohajer is with the Department of Electrical Engineering and Computer Science, University of California, Berkeley,
CA, USA. E-mail: mohajer@eecs.berkeley.edu.}
\thanks{ H. Vincent Poor is with the Department of Electrical Engineering, Princeton University, Princeton,
NJ, USA. E-mail: poor@princeton.edu.}%
\thanks{ Shlomo Shamai is with the Department of Electrical Engineering, Technion, Israel Institute of Technology, Haifa, Israel. 
E-mail: sshlomo@ee.technion.ac.il.}%
\thanks{The work of H. V. Poor was supported in part by the Air Force Office of Scientific Research under MURI Grant FA 9550-09-1-0643. The work of S. Shamai was supported by the Israel Science Foundation (ISF), and the Philipson Fund for Electrical Power. This paper was presented in part at the IEEE International Conference on Communications (ICC) 2012, Ottawa, Canada. This work was done while R. Tandon and S. Mohajer were with Princeton University.}%
\thanks{Copyright (c) 2012 IEEE. Personal use of this material is permitted.  However, permission to use this material for any other purposes must be obtained from the IEEE by sending a request to pubs-permissions@ieee.org.}}

\markboth{IEEE Transactions on Information Theory,~Vol.~x, No.x~, Month~20xx}%
{Shell \MakeLowercase{\textit{et al.}}: Bare Demo of IEEEtran.cls for Journals}
\maketitle

\newcommand{\DoF}{\mbox{DoF}}
\newtheorem{Theo}{Theorem}
\newtheorem{Rem}{Remark}
\newtheorem{Lem}{Lemma}
\newtheorem{Cor}{Corollary}
\newtheorem{Def}{Definition}
\newtheorem{Prop}{Proposition}
\newtheorem{claim}{Claim}

\begin{abstract}
The two-user multiple-input multiple-output (MIMO) interference channel (IC) with arbitrary number of antennas at each terminal is considered and the degrees of freedom ($\DoF$) region is characterized in the presence of noiseless channel output feedback from each receiver to its respective transmitter and availability of delayed channel state information at the transmitters (CSIT). It is shown that having output feedback and delayed CSIT can strictly enlarge the $\DoF$ region of the MIMO IC when compared to the case in which only delayed CSIT is present. The proposed coding schemes that achieve the corresponding $\DoF$ region with feedback and delayed CSIT utilize both resources, i.e., feedback and delayed CSIT in a non-trivial manner. It is also shown that the $\DoF$ region with local feedback and delayed CSIT is equal to the $\DoF$ region with global feedback and delayed CSIT, i.e., local feedback and delayed CSIT is equivalent to global feedback and delayed CSIT from the perspective of the degrees of freedom region.  The converse is proved for a stronger setting in which the channels to the two receivers need not be statistically equivalent. 
\end{abstract}


\IEEEpeerreviewmaketitle

\section{Introduction}
In many wireless networks, multiple pairs of transmitters/receivers wish to communicate over a shared medium. In such situations, due to the broadcast and superposition nature of the wireless medium, the effect of interference is inevitable. Hence, management of interference is of extreme importance in such networks. Various interference management techniques have been proposed over the past few decades. The more traditional approaches to deal with interference either treat it as noise (in the low interference regime) or decode and then remove it from the received signal (in the high interference regime). However, such techniques are not strong enough to achieve the optimal performance of the network even in the simple interference channel with two pairs of multiple-input multiple-output (MIMO) transceivers. Recently, more sophisticated schemes, such as interference alignment \cite{Jafar:alignmenet, MaddahAli:X} and (aligned) interference neutralization \cite{mohajer:neutralization,Jafar:neutralization} have been proposed for managing interference, which can significantly increase the achievable rate over the interference networks (also see \cite{Jafar:Tutorial} for an excellent tutorial on interference alignment). However, these techniques are usually based on availability of instantaneous (perfect) channel state information at the transmitters (p-CSIT). Such an assumption is perhaps not very realistic in practical systems, at least when dealing with fast fading links.

Quite surprisingly, it is shown by Maddah-Ali and Tse \cite{MaddahAli-Tse:DCSI-BC} that even delayed (stale) CSIT is helpful to improve the achievable rate of wireless network with multiple flows, even if the channel realizations vary independently across time. In  \cite{MaddahAli-Tse:DCSI-BC}, the authors studied a  two user multiple-input single-output (MISO) broadcast channel (BC) with two transmit antennas and one antenna at each receiver, where the channels between the transmitter and receivers change over time from one channel use to the next independently, and channel state information is available to the transmitters only at the end of each channel use. They showed that the sum degrees of freedom ($\DoF$) of $4/3$ is achievable for this network, which is in contrast to $\DoF=1$, which is known to be optimal for the case of no CSIT.  This result is also extended in \cite{MaddahAli-Tse:DCSI-BC} to the $K$ user MISO BC, and extensions to certain MIMO BCs have been reported in \cite{Vaze:MIMOBC}. 

This usefulness of delayed CSIT for interference networks is further explored in \cite{Retro-IA}, where it is shown that for the single input single output (SISO) three user interference channel (IC) and the two user X-channel, $9/8$ and $8/7$ $\DoF$ are achievable with delayed CSIT respectively. The $\DoF$ region of the two-user MIMO interference channel with delayed CSIT is completely characterized by Vaze and Varanasi \cite{Vaze:MIMOICDelayedCSIT}. It is shown that, depending on the number of antennas at each terminal, the $\DoF$ with delayed CSIT can be strictly better than that of no CSIT \cite{Vaze:NoCSIT-A,Vaze:NoCSIT-B}, and worse than that with instantaneous CSIT \cite{JafarFakhereddin:MIMOIC}. 

The role of output feedback in the performance of wireless communication systems has received considerable attention over the past few decades. It is well known that feedback does not increase the capacity of point-to-point discrete memoryless channels \cite{zero:56}. Unlike the point-to-point case, feedback can increase the capacity of the multiple-access channel \cite{GW:1975} and broadcast channel \cite{Ozarow:BCFB}. The effects of feedback on the capacity region of the static (non-fading) interference channel have been studied in several recent papers (see \cite{SuhTseIT}, \cite{prabhakaran2011interference} and references therein). In particular, the approximate capacity of the $2$-user symmetric Gaussian interference channel is obtained in \cite{prabhakaran2011interference}, where it is shown that simple coding schemes achieve within a constant gap from the symmetric capacity in the presence of noiseless feedback. The entire approximate feedback capacity region of the  $2$-user Gaussian interference channel with arbitrary channel gains has been characterized independently in \cite{SuhTseIT}, which reveals that output feedback strictly improves the number of generalized degrees of freedom of the interference channel. These works show that in the case of static channels, presence of output feedback not only has impact on the capacity of the channel (e.g. in the multiple access and broadcast channels), but can also enlarge the number of generalized degrees of freedom. 

On the other hand, the usefulness of output feedback for the fast fading interference and X-channels with no CSIT was presented in \cite{Retro-IA}. It is shown in \cite{Retro-IA} that $6/5$ and $4/3$ $\DoF$ are achievable for the three user SISO IC and the two user SISO X-channel respectively with output feedback. An immediate  consequence is that output feedback is beneficial when no channel state information is available at the transmitters. One question that can be raised here is whether output feedback can be helpful with delayed CSIT or not. For the case of the broadcast channel (BC), this question is answered in a negative way in \cite{Vaze:MIMOBC}: having output feedback in addition to delayed CSIT does not increase the $\DoF$ region of the MIMO BC. 

In this work, we study this question for the two-user MIMO IC, in which each transmitter is provided with the past state information of the channel (i.e., with delayed CSIT), as well as the received signal (output feedback) from its respective receiver.  It turns out that existence of output feedback can increase the $\DoF$ region of the interference channel. This is indeed surprising as it has been shown in \cite{CadambeJafar2009:cooperation} and \cite{HuangJafar2009} that with perfect (instantaneous) CSIT, any form of cooperation (which may include feedback) does not increase the $\DoF$ region of the MIMO interference channel under the time-varying/frequency-selective channel model.
 The benefit of output feedback with delayed CSIT becomes apparent from the following observation:  in the presence of output feedback with delayed channel state information, Transmitter $1$ in addition to being able to reconstruct the interference it caused at Receiver $2$, can also reconstruct a part of the signal intended to Receiver $2$. This is in contrast to the case of the MIMO BC, where all information symbols are created at one transmitter, and hence output feedback in addition to delayed CSIT does not increase the $\DoF$ region of the MIMO BC, even though it may increase the capacity region. 

The main contribution of this paper is the characterization of the $\DoF$ region of the two user MIMO IC with local output feedback and delayed CSIT. It is also shown that this $\DoF$ region remains the same even if global feedback is present from both receivers to both transmitters. That is, local feedback and delayed CSIT is equivalent to the enhanced setting of global feedback and delayed CSIT from the perspective of the degrees of freedom region. We note here that the same set of results have also been reported independently in \cite{Vaze:ShannonFB}. Our results are stronger in the sense that we do not assume the channels at the two receivers to be statistically equivalent. Typically most of the converse proofs for delayed CSIT scenarios are based on the assumption in which the channels to different receivers are generated from identical distributions. Recently, a novel approach was presented in  \cite{GouJafar:MixedCSIT}, in which the restrictive statistical equivalence assumption was relaxed and a strengthened  converse was proved for the two user MISO broadcast channel with delayed CSIT. Our converse approach is inspired by the approach taken in \cite{GouJafar:MixedCSIT}.  Furthermore, from our converse proof we find an interesting connection of the MIMO IC with feedback and delayed CSIT to a physically degraded cognitive MIMO IC with no CSIT. Parts of this work have been presented in \cite{ICC2012Tandon}.

\section{MIMO IC with Feedback and Delayed CSIT}
We consider the $(M_{1},M_{2},N_{1},N_{2})$ MIMO-IC for which the transmitters are denoted as $\mathbf{Tx}_{1}$ and $\mathbf{Tx}_{2}$; and the receivers as $\mathbf{Rx}_{1}$ and $\mathbf{Rx}_{2}$.
$\mathbf{Tx}_{m}$ intends to send a message $W_{m}$ to $\mathbf{Rx}_{m}$, for $m=1,2$ and the messages $W_{1}$ and $W_{2}$ are independent.
The channel outputs at the Receivers are given as
\begin{align}
Y_{1}(i)&= \mathbf{H}_{11}(i)X_{1}(i)+ \mathbf{H}_{12}(i)X_{2}(i)+ Z_{1}(i)\nonumber\\
Y_{2}(i)&= \mathbf{H}_{21}(i)X_{1}(i)+ \mathbf{H}_{22}(i)X_{2}(i)+ Z_{2}(i)\nonumber,
\end{align}
where $X_{m}(i)$ is the signal transmitted by the $m$th transmitter $\mathbf{Tx}_{m}$; $\mathbf{H}_{nm}(i)\in \mathbb{C}^{N_{n}\times M_{m}}$
denotes the channel matrix between the $n$th receiver and $m$th transmitter; and $Z_{n}(i)\sim \mathcal{CN}(0,I_{N_{n}})$, for $n=1,2$,
is additive noise at Receiver $n$. The power constraints are $\mathbb{E}||X_{m}(i)||^{2}\leq P$, for $\forall$  $m,i$.

We make the following assumptions for the channel matrices:
\begin{itemize}
\item A1: All elements of $\mathbf{H}_{nm}(i)$ are independently and identically distributed (i.i.d.) from a continuous distribution, which we symbolically denote by $\lambda_{nm}$.
\item A2: The distributions $\lambda_{11}, \lambda_{12}, \lambda_{21}, \lambda_{22}$ are {\emph{not}} necessarily identical. 
\item A3: The channel matrices vary in an i.i.d. manner across time.
\end{itemize}
 
We denote by $\mathbf{H}(i)= \{\mathbf{H}_{11}(i),\mathbf{H}_{12}(i),\mathbf{H}_{21}(i),\mathbf{H}_{22}(i)\}$ the collection of all channel matrices at time $i$. Furthermore, $\mathbf{H}^{i-1}=\{\mathbf{H}(1),\mathbf{H}(2),\ldots,\mathbf{H}(i-1)\}$ denotes the set of all channel matrices up till time $(i-1)$. Similarly, we denote by $Y_{n}^{i-1}=\{Y_{n}(1),\ldots,Y_{n}(i-1)\}$ the set of all channel outputs at Receiver $n$ up till time $(i-1)$.
\begin{figure}[t]
\centering \includegraphics[width=6.2cm, height=6.2cm]{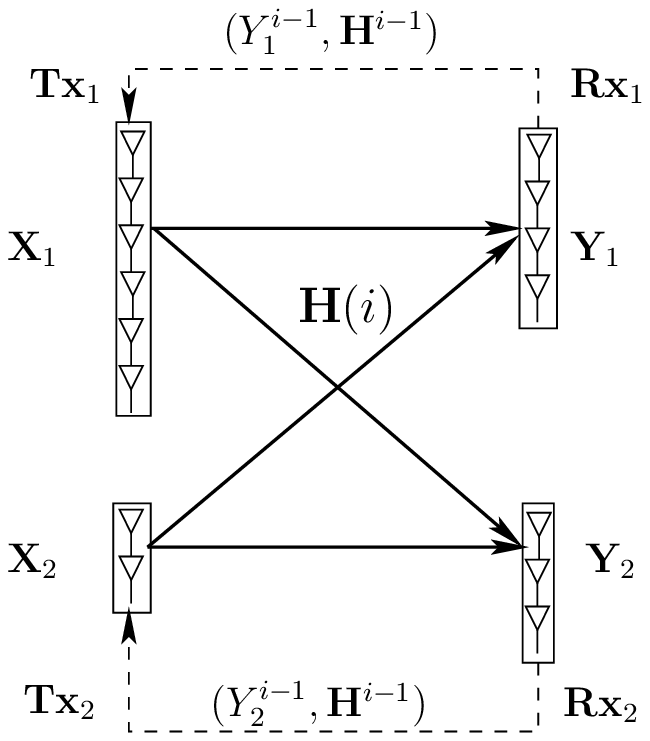}
\caption{The MIMO-IC with local output feedback and delayed CSIT.}\label{Figmodel}
\vspace{-0.7cm}
\end{figure}

We assume that both receivers have the knowledge of global and instantaneous channel state information, i.e., both receivers have access to $\mathbf{H}^i=\{\mathbf{H}(1),\mathbf{H}(2),\ldots,\mathbf{H}(i)\}$ at time $i$, for all $i$.  Depending on the availability of the amount of feedback and channel state information at the transmitters, permissible encoding functions and the corresponding DoF regions can be different. Below, we enumerate the permissible encoding functions and the corresponding DoF regions for several scenarios as follows:
\begin{itemize}
\item No CSIT: $\DoF^{\mathrm{No-CSIT}}$
\begin{align}
 X_{m}(i)=f_{m,i}(W_{m}), \quad m=1,2.\nonumber
\end{align}
\item Perfect CSIT: $\DoF^{\mathrm{p-CSIT}}$
\begin{align}
 X_{m}(i)=f_{m,i}(W_{m},\mathbf{H}^{i}), \quad m=1,2.\nonumber
\end{align}
\item Delayed CSIT: $\DoF^{\mathrm{d-CSIT}}$
\begin{align}
X_{m}(i)=f_{m,i}(W_{m},\mathbf{H}^{i-1}), \quad m=1,2.\nonumber
\end{align}
\item Local output feedback: $\DoF^{\mathrm{FB}}$
\begin{align}
X_{m}(i)=f_{m,i}(W_{m},Y_{m}^{i-1}), \quad m=1,2.\nonumber
\end{align}
\item Local output feedback and delayed CSIT: $\DoF^{\mathrm{FB,d-CSIT}}$
\begin{align}
X_{m}(i)=f_{m,i}(W_{m},Y_{m}^{i-1},\mathbf{H}^{i-1}), \quad m=1,2.\nonumber
\end{align}
\item Global output feedback and delayed CSIT: $\DoF^{\mathrm{GFB,d-CSIT}}$
\begin{align}
X_{m}(i)=f_{m,i}(W_{m},Y_{1}^{i-1}, Y_{2}^{i-1}, \mathbf{H}^{i-1}), \quad m=1,2.\nonumber
\end{align}
\end{itemize}

A coding scheme with block length $n$ for the MIMO-IC for a given feedback and CSIT configuration\footnote{A configuration could correspond to either one of the following scenarios: No CSIT, Perfect CSIT, Delayed CSIT, Local output feedback, Local output feedback and delayed CSIT, or  Global output feedback and delayed CSIT.} consists of a sequence of encoding functions $\{X_{m}(i)\}_{i=1}^{n}$, for $m=1,2$
and two decoding functions
\begin{align}
\hat{W}_{1}= g^{n}_{1}(Y_{1}^{n},\mathbf{H}^{n}),\quad
\hat{W}_{2}= g^{n}_{2}(Y_{2}^{n},\mathbf{H}^{n}).\nonumber
\end{align}
A rate pair $(R_{1}(P),R_{2}(P))$ is achievable if there exists a sequence of coding schemes such that $\mathbb{P}(W_{m}\neq \hat{W}_{m})\rightarrow 0$
as $n\rightarrow \infty$ for both $m=1,2$. The capacity region $\mathcal{C}(P)$ is defined as the closure of the set of all achievable rate pairs $(R_{1}(P),R_{2}(P))$.
We define the DoF region as follows:
\begin{align}
\DoF&=\Big\{(d_{1},d_{2})\bigg| d_{m}\geq 0, \mbox{ and } 
 \exists (R_{1}(P),R_{2}(P))\in \mathcal{C}(P) \nonumber\\
&\hspace{2cm} \mbox{ s.t. } d_{m}=\lim_{P\rightarrow \infty}\frac{R_{m}(P)}{\log_{2}(P)}, m=1,2 \Big\}.\nonumber
\end{align}

\section{Main Results and Discussion}
The main contribution of this paper is a complete characterization of $\DoF^{\mathrm{FB,d-CSIT}}$ \break and $\DoF^{\mathrm{GFB,d-CSIT}}$, stated in the following theorem:
\begin{Theo}\label{Theorem1}
The $\DoF$ region of the two user MIMO IC with local feedback and delayed CSIT is equal to the $\DoF$ region with 
global feedback and delayed CSIT, i.e., 
\begin{align}
\DoF^{\mathrm{GFB,d-CSIT}}&=\DoF^{\mathrm{FB,d-CSIT}},
\end{align}
and this region is given by the set of all non-negative pairs $(d_{1},d_{2})$ that satisfy
\begin{align}
&\quad \hspace{0.17in}d_{1}\leq \min(M_{1},N_{1})\label{SU1}\\
&\quad \hspace{0.17in}d_{2}\leq \min(M_{2},N_{2})\label{SU2}\\
&d_{1}+d_{2}\leq \min\Big\{M_{1}+M_{2}, N_{1}+N_{2},\nonumber\\ &\hspace{2.5cm}\max(M_{1},N_{2}),
\max(M_{2},N_{1})\Big\}\label{Sum}\\
&\frac{d_{1}}{\min(N_{1}+N_{2},M_{1})}+\frac{d_{2}}{\min(N_{2},M_{1})}\leq \frac{\min(N_{2},M_{1}+M_{2})}{\min(N_{2},M_{1})}\label{L1}\\
&\frac{d_{1}}{\min(N_{1},M_{2})}+\frac{d_{2}}{\min(N_{1}+N_{2},M_{2})}\leq \frac{\min(N_{1},M_{1}+M_{2})}{\min(N_{1},M_{2})}\label{L2}.
\end{align}
\end{Theo}

For comparison, we recall the DoF region with perfect, instantaneous CSIT at the transmitters  $\DoF^{\mathrm{p-CSIT}}$ \cite{JafarFakhereddin:MIMOIC}:
\begin{align}
d_{1}&\leq \min(M_{1},N_{1})\label{PSU1}\\
d_{2}&\leq \min(M_{2},N_{2})\label{PSU2}\\
d_{1}+d_{2}&\leq \min\Big\{M_{1}+M_{2}, N_{1}+N_{2},\nonumber\\ &\hspace{1.3cm}\max(M_{1},N_{2}),
\max(M_{2},N_{1})\Big\}.\label{PSum}
\end{align}
In addition, the $\DoF$ region with delayed CSIT, $\DoF^{\mathrm{d-CSIT}}$ was characterized in \cite{Vaze:MIMOICDelayedCSIT}. This region is given by the
set of inequalities as in Theorem \ref{Theorem1} along with two more inequalities. In particular, to characterize $\DoF^{\mathrm{d-CSIT}}$, \cite{Vaze:MIMOICDelayedCSIT} defines 
two mutually exclusive conditions:
\begin{align}
&\mbox{Condition } 1: \nonumber\\
&M_{1}>N_{1}+N_{2}-M_{2}>N_{1}>N_{2}>M_{2}>N_{2}\left(\frac{N_{2}-M_{2}}{N_{1}-M_{2}}\right)\label{eq:dcsi-cond-1}\\
&\mbox{Condition } 2: \nonumber\\
&M_{2}>N_{1}+N_{2}-M_{1}>N_{2}>N_{1}>M_{1}>N_{1}\left(\frac{N_{1}-M_{1}}{N_{2}-M_{1}}\right). \label{eq:dcsi-cond-2}
\end{align}
If condition $1$ holds, then $\DoF^{\mathrm{d-CSIT}}$ is given by the inequalities in Theorem \ref{Theorem1} and
the following additional bound (bound $L_{4}$ in \cite{Vaze:MIMOICDelayedCSIT}):
\begin{align}
d_{1}+d_{2}\left(\frac{N_{1}+2N_{2}-M_{2}}{N_{2}}\right)\leq N_{1}+N_{2}.
\label{eq:dcsi-1}
\end{align}
If condition $2$ holds, then $\DoF^{\mathrm{d-CSIT}}$ is given by the inequalities in Theorem \ref{Theorem1} and
the following additional bound (bound $L_{5}$ in \cite{Vaze:MIMOICDelayedCSIT}):
\begin{align}
d_{2}+d_{1}\left(\frac{N_{2}+2N_{1}-M_{1}}{N_{1}}\right)\leq N_{1}+N_{2}.
\label{eq:dcsi-2}
\end{align}

Hence, from Theorem \ref{Theorem1}, we have the following relationship:
\begin{align}
\DoF^{\mathrm{No-CSIT}}\subseteq \DoF^{\mathrm{d-CSIT}}\subseteq \DoF^{\mathrm{FB,d-CSIT}}\subseteq \DoF^{\mathrm{p-CSIT}}.\nonumber
\end{align}

Converse proof for Theorem \ref{Theorem1}: the upper bounds (\ref{SU1})-(\ref{SU2}) are straightforward from the point-to-point MIMO channel.
It has been shown in \cite{CadambeJafar2009:cooperation} and \cite{HuangJafar2009} that any form of feedback/cooperation does not increase $\DoF^{\mathrm{p-CSIT}}$. 
Therefore, $\DoF^{\mathrm{p-CSIT}}$ is an outer bound on the $\DoF$ region with global feedback and delayed CSIT.
Hence, to show that $\DoF^{\mathrm{GFB,d-CSIT}}$ is contained in the region given by (\ref{SU1})-(\ref{L2}), we need only to prove the bounds
(\ref{L1}) and (\ref{L2}) for the case of global feedback and delayed CSIT. Since (\ref{L1}) and (\ref{L2}) are symmetric, we need to prove that
if $(d_{1},d_{2})\in \DoF^{\mathrm{FB,d-CSIT}}$, then $(d_{1},d_{2})$ must satisfy the bound (\ref{L1}). 
We establish the bound (\ref{L1}) in the Section~\ref{sec:L1}.

Coding schemes with feedback and delayed CSIT that achieve the region $\DoF^{\mathrm{FB,d-CSIT}}$ stated in Theorem \ref{Theorem1} are presented in Section \ref{Section:codingMIMO}.

\section{Coding for MIMO IC with Local Feedback and Delayed CSIT}\label{Section:codingMIMO}
In this section, we present coding schemes that achieve the $\DoF$ region for the MIMO IC stated in Theorem \ref{Theorem1}.
We assume without loss of generality, that $N_{1}\geq N_{2}$. We refer the reader to Table I in reference \cite{Vaze:MIMOICDelayedCSIT}.
\begin{itemize}
\item If $(M_{1},M_{2},N_{1},N_{2})$ are such that
\begin{align}
\DoF^{\mathrm{FB,d-CSIT}}=\DoF^{\mathrm{d-CSIT}}\label{equality},
\end{align}
 coding schemes presented in \cite{Vaze:MIMOICDelayedCSIT} which use delayed CSIT only suffice for our problem.
The condition (\ref{equality}) corresponds to cases A.I, A.II, B.$0$, B.I, and B.II, as defined in \cite{Vaze:MIMOICDelayedCSIT}.
\item If $(M_{1},M_{2},N_{1},N_{2})$ are such that
\begin{align}
\DoF^{\mathrm{FB,d-CSIT}}\supset\DoF^{\mathrm{d-CSIT}},
\end{align}
we present a novel coding scheme that achieves $\DoF^{\mathrm{FB,d-CSIT}}$.
\end{itemize}
We present the optimal coding schemes for the case of arbitrary $(M_{1},M_{2},N_{1},N_{2})$-MIMO IC in Section \ref{Section:codingmimogeneral}. In the following sub-sections, we highlight the contribution of our coding scheme through two examples that capture its essential features and lead to valuable insights for the case of the general $(M_{1},M_{2},N_{1},N_{2})$-MIMO IC.

\subsection{$(6,2,4,3)$-IC with Feedback and Delayed CSIT}
\begin{figure}[t]
\centering \includegraphics[width=9.0cm, height=5.3cm]{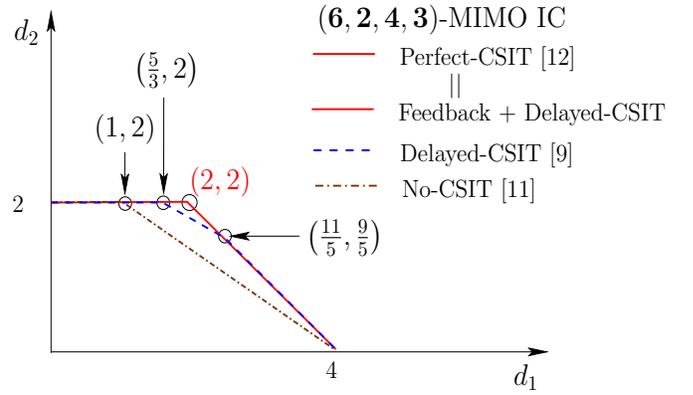}
  \caption{$\DoF$ region for the $(6,2,4,3)$-MIMO-IC with various assumptions.}\label{Figexample1}
\end{figure}

We first focus on the case of the $(6,2,4,3)$-MIMO IC. For comparison purposes, we note here the $\DoF$ regions with no-CSIT, perfect CSIT,
delayed CSIT, output feedback and delayed CSIT. For all these four regions, we have the following bounds:
\begin{align}
d_{1}\leq 4;\quad d_{2}\leq 2.\nonumber
\end{align}
Besides these, we have the following additional bounds:
\begin{itemize}
\item No-CSIT:
\begin{align}
d_1 + \frac{3}{2} d_2 \leq 4. \nonumber
\end{align}
\item Perfect CSIT:
\begin{align}
d_{1}+d_{2}&\leq 4.\nonumber
\end{align}
\item Delayed CSIT (Case B-III, \cite{Vaze:MIMOICDelayedCSIT}):
\begin{align}
d_{1}+d_{2}\leq 4;\quad  d_{1}+\frac{8d_{2}}{3}\leq 7.\nonumber
\end{align}
\item Output feedback and delayed CSIT (Theorem \ref{Theorem1}):
\begin{align}
d_{1}+d_{2}\leq 4;\quad \frac{d_{1}}{6}+\frac{d_{2}}{3}\leq 1.\nonumber
\end{align}
\end{itemize}
It can be verified that this region is the same as the DoF region with perfect CSIT, since the bound $\frac{d_{1}}{6}+\frac{d_{2}}{3}\leq 1$
is redundant as $d_{1}\geq 2\geq d_{2}$ is a valid choice (see Figure \ref{Figexample1}).

The main contribution of the coding scheme is to show the achievability of the point $(2,2)$ under the assumption of output feedback and delayed CSIT.
To show the achievability of point $(2,2)$, we will show that in three uses of the channel, we can reliably
transmit $6$ information symbols to Receiver $1$, and $6$ information symbols to Receiver $2$.

\begin{figure*}[t]
\centering \includegraphics[width=17.5cm]{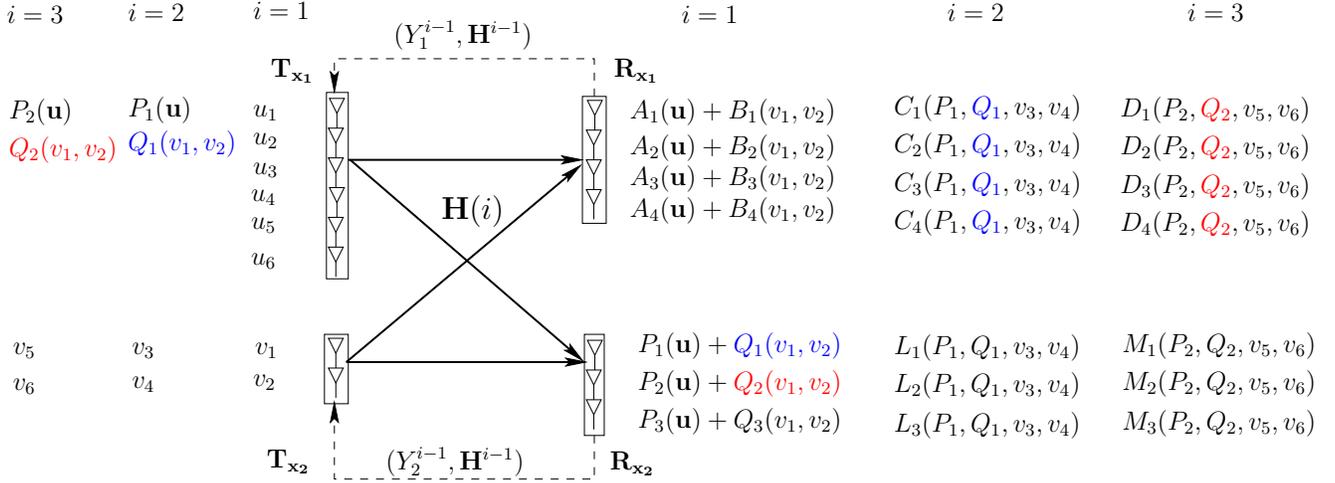}
  \caption{Coding scheme with FB and delayed CSIT: $(6,2,4,3)$-MIMO-IC.}\label{Figcodingexample1}
\end{figure*}
Encoding at Transmitter $2$: Transmitter $2$ sends fresh information symbols on both its antennas for $i=1,2,3$, i.e., the channel input of Transmitter $2$, denoted as $X_{2}(i)$ for $i=1,2,3$, can be written as
\begin{align}
X_{2}(1)&=[v_{1} \hspace{0.1in} v_{2}]^{T}, X_{2}(2)=[v_{3} \hspace{0.1in} v_{4}]^{T}, X_{2}(3)=[v_{5} \hspace{0.1in} v_{6}]^{T}.\nonumber
\end{align}
At $i=1$, Transmitter $1$ sends $6$ information symbols on its $6$ antennas, i.e., it sends
\begin{align}
X_{1}(1)&=[u_{1} \hspace{0.1in} u_{2} \hspace{0.1in} u_{3} \hspace{0.1in}u_{4} \hspace{0.1in}u_{5} \hspace{0.1in}u_{6}]^{T}.
\end{align}
Let us denote by $\mathbf{u}=(u_{1},\ldots,u_{6})$ the vector of information symbols intended for Receiver $1$.
The outputs at Receivers $1$ and $2$ at $i=1$ (ignoring noise) are given as
\begin{align}
Y_{1}(1)&=\left[
\begin{array}{c}
  A_{1}(\mathbf{u})+B_{1}(v_{1},v_{2}) \\
  A_{2}(\mathbf{u})+B_{2}(v_{1},v_{2}) \\
  A_{3}(\mathbf{u})+B_{3}(v_{1},v_{2}) \\
  A_{4}(\mathbf{u})+B_{4}(v_{1},v_{2})
\end{array}\right],\nonumber
\end{align}
\begin{align}
Y_{2}(1)&=\left[
\begin{array}{c}
  P_{1}(\mathbf{u})+Q_{1}(v_{1},v_{2}) \\
  P_{2}(\mathbf{u})+Q_{2}(v_{1},v_{2}) \\
  P_{3}(\mathbf{u})+Q_{3}(v_{1},v_{2})
\end{array}\right].\nonumber
\end{align}

Upon receiving $Y_{1}(i)$ (channel output feedback) from Receiver $1$ and $H(1)$ (delayed CSIT), Transmitter $1$ can use $(u_{1},\ldots,u_{6},Y_{1}(1), H(1))$ to solve for $(v_{1},v_{2})$. Consequently, it can reconstruct $Q_{1}(v_{1},v_{2})$ and $Q_{2}(v_{1},v_{2})$ which constitute a part of the received signal, $Y_{2}(1)$, at Receiver $2$. In addition, having access to delayed CSIT, $H(1)$, it can also compute $P_{1}(\mathbf{u})$ and $P_{2}(\mathbf{u})$, a part of the interference it caused at Receiver $2$. In the next two time instants, i.e., at $i=2$ and $3$, Transmitter $1$ sends
\begin{align}
X_{1}(2)&=[P_{1}(\mathbf{u}) \hspace{0.1in} Q_{1}(v_{1},v_{2}) \hspace{0.1in} \phi \hspace{0.1in}\phi \hspace{0.1in}\phi \hspace{0.1in}\phi]^{T}\\
X_{1}(3)&=[P_{2}(\mathbf{u}) \hspace{0.1in} Q_{2}(v_{1},v_{2}) \hspace{0.1in} \phi \hspace{0.1in}\phi \hspace{0.1in}\phi \hspace{0.1in}\phi]^{T},
\end{align}
where $\phi$ denotes a constant symbol known to all terminals.

The channel outputs at Receiver $1$ at $i=2,3$ are given as follows:
\begin{align}
Y_{1}(2)&=\left[
\begin{array}{c}
  C_{1}(P_{1}(\mathbf{u}),Q_{1}(v_{1},v_{2}),v_{3},v_{4})\\
  C_{2}(P_{1}(\mathbf{u}),Q_{1}(v_{1},v_{2}),v_{3},v_{4})\\
  C_{3}(P_{1}(\mathbf{u}),Q_{1}(v_{1},v_{2}),v_{3},v_{4})\\
  C_{4}(P_{1}(\mathbf{u}),Q_{1}(v_{1},v_{2}),v_{3},v_{4})
  \end{array}\right],\nonumber\\
Y_{1}(3)&=\left[
\begin{array}{c}
  D_{1}(P_{2}(\mathbf{u}),Q_{2}(v_{1},v_{2}),v_{5},v_{6})\\
  D_{2}(P_{2}(\mathbf{u}),Q_{2}(v_{1},v_{2}),v_{5},v_{6})\\
  D_{3}(P_{2}(\mathbf{u}),Q_{2}(v_{1},v_{2}),v_{5},v_{6})\\
  D_{4}(P_{2}(\mathbf{u}),Q_{2}(v_{1},v_{2}),v_{5},v_{6})
  \end{array}\right].\nonumber
  \end{align}

Decoding at Receiver $1$: Note that $Y_{1}(2)$ is comprised of $4$ equations in $4$  variables $(P_{1}(\mathbf{u}),Q_{1}(v_{1},v_{2}),v_{3},v_{4})$, whose coefficients are drawn from a continuous distribution. Therefore, they are linearly independent almost surely, and the coefficient matrix is full-rank. Hence, Receiver $1$ can decode  $4$ symbols $(P_{1}(\mathbf{u}), Q_{1}(v_{1},v_{2}), v_{3},v_{4})$ by matrix inversion. Similarly, $Y_{1}(3)$ is comprised of $4$ (almost surely) linearly independent equations in $4$ variables $(P_{2}(\mathbf{u}),Q_{2}(v_{1},v_{2}),v_{5},v_{6})$. Hence Receiver $1$ can decode $(P_{2}(\mathbf{u}), Q_{2}(v_{1},v_{2}), v_{5},v_{6})$ from $Y_{1}(3)$. Having decoded $(Q_{1}(v_{1},v_{2}), Q_{2}(v_{1},v_{2}))$, it can solve for $(v_{1},v_{2})$ and compute $\{B_{j}(v_{1},v_{2})\}_{j=1}^{4}$, the interference caused at $i=1$. Upon subtracting $B_{j}(v_{1},v_{2})$ from the output at the $j$th antenna corresponding to $Y_{1}(1)$, Receiver $1$ obtains  $(A_{1}(\mathbf{u}),A_{2}(\mathbf{u}),A_{3}(\mathbf{u}),A_{4}(\mathbf{u}), P_{1}(\mathbf{u}),P_{2}(\mathbf{u}))$, i.e., it has $6$ linearly independent equations in $6$ variables $(u_{1},\ldots,u_{6})$. Hence all $6$ information symbols $(u_{1},\ldots,u_{6})$ can be decoded by Receiver $1$ in three uses of the channel.

The channel outputs at Receiver $2$ at $i=2,3$ are given as follows:
\begin{align}
Y_{2}(2)&=\left[
\begin{array}{c}
  L_{1}(P_{1}(\mathbf{u}),Q_{1}(v_{1},v_{2}),v_{3},v_{4})\\
  L_{2}(P_{1}(\mathbf{u}),Q_{1}(v_{1},v_{2}),v_{3},v_{4})\\
  L_{3}(P_{1}(\mathbf{u}),Q_{1}(v_{1},v_{2}),v_{3},v_{4})
\end{array}\right],\nonumber\\
Y_{2}(3)&=\left[
\begin{array}{c}
  M_{1}(P_{2}(\mathbf{u}),Q_{2}(v_{1},v_{2}),v_{5},v_{6})\\
  M_{2}(P_{2}(\mathbf{u}),Q_{2}(v_{1},v_{2}),v_{5},v_{6})\\
  M_{3}(P_{2}(\mathbf{u}),Q_{2}(v_{1},v_{2}),v_{5},v_{6})
  \end{array}\right].\nonumber
  \end{align}

Decoding at Receiver $2$: at Receiver $2$, we have $9$ linearly independent equations (from $Y_{2}(1),Y_{2}(2),Y_{2}(3)$) in $9$ variables $(v_{1},\ldots,v_{6},P_{1}(\mathbf{u}),\ldots,P_{3}(\mathbf{u}))$, where $(P_{1}(\mathbf{u}),\ldots,P_{3}(\mathbf{u}))$ is the additive interference
at $i=1$. Hence it can decode $6$ information symbols \break $(v_{1},v_{2},v_{3},v_{4},v_{5},v_{6})$ in three uses of the channel.
Hence, we have shown the achievability of the point $(2,2)$ with channel output feedback and delayed CSIT.

{\bf{Remark $1$}}: It is instructive to compare this coding scheme to the case of delayed CSIT. In particular, the point $(5/3,2)$ lies on the
boundary of $\DoF^{\mathrm{d-CSIT}}$. In the coding scheme that achieves this point, it suffices to transmit $5$ symbols to Receiver $1$
and $6$ symbols to Receiver $2$ in three channel uses. Under the delayed CSIT assumption, Transmitter $1$ can at best reconstruct the interference it
caused at Receiver $2$. In the terminology of the coding scheme described above, Transmitter $1$ can reconstruct $(P_{1}(\mathbf{u}),P_{2}(\mathbf{u}),P_{3}(\mathbf{u}))$. In subsequent channel uses,  Transmitter $1$ sends $P_{1}(\mathbf{u})$ and $P_{2}(\mathbf{u})$ in $i=2$, and 
$P_{3}(\mathbf{u})$ in $i=3$.  At the end of transmission, note that Receiver $2$ still has $9$ equations in $9$ variables, $(v_{1},\ldots,v_{6},P_{1}(\mathbf{u}),\ldots,P_{3}(\mathbf{u}))$, therefore it can reliably decode $(v_{1},\ldots,v_{6})$.

The difference between the optimal coding schemes for these two models is highlighted by the decoding capability of Receiver $1$. For instance, in the scheme with delayed CSIT alone, Receiver $1$ has $11$ linearly independent equations ($4$ equations for $i=1$, $4$ equations in $i=2$, and $3$ equations in $i=3$) in $12$ variables $(u_{1},\ldots,u_{6},v_{1},\dots,v_{6})$; hence at best it can decode any $5$ of the $6$ information symbols $(u_{1},\ldots,u_{6})$.  On the other hand, in our scheme, which allows for output feedback along with delayed CSIT, Transmitter $1$ can exactly separate the interference and signal component of Receiver $2$, i.e., besides knowing $(P_{1}(\mathbf{u}),\ldots,P_{3}(\mathbf{u}))$, it can also exactly reconstruct $(Q_{1}(v_{1},v_{2}),Q_{2}(v_{1},v_{2}))$ (see Figure \ref{Figcodingexample1}, which also highlights the difference of the coding schemes). This additional knowledge of $(Q_{1}(v_{1},v_{2}),Q_{2}(v_{1},v_{2}))$ is useful in transmission of one additional symbol to Receiver $1$ in three channel uses.

{\bf{Remark $2$}}: We note here that for this particular example, we can achieve the $\DoF$ region with perfect instantaneous CSIT.
Recall from \cite{HuangJafar2009} that the point $(2,2)$ can be achieved with perfect CSIT in one shot, i.e., in one channel use. As we have shown,
output feedback and delayed CSIT can also achieve the point $(2,2)$, albeit, we pay the price of a larger delay, i.e., we can achieve this
pair in three channel uses. Moreover, it is clear that the role of feedback and delayed CSIT is crucial in the proposed coding scheme, and it is not possible to achieve the same point in a single shot. This observation also highlights the delay penalty incurred by the {\emph{causal} knowledge of output feedback and delayed CSIT.

From this example it is clear that $\DoF^{\mathrm{FB,d-CSIT}}=\DoF^{\mathrm{p-CSIT}}$. However, this equality does not hold in general.
In the next section, we illustrate this by an example for which $\DoF^{\mathrm{d-CSIT}}\subset\DoF^{\mathrm{FB,d-CSIT}}\subset\DoF^{\mathrm{p-CSIT}}$, i.e., having feedback and delayed CSIT is strictly worse than having perfect CSIT and strictly better than having only delayed CSIT.

\subsection{$(8,4,6,5)$-MIMO IC}
We now focus on the $(8,4,6,5)$-MIMO IC (see Figure \ref{Figexample2}).
The main contribution is to show the achievability of the point $(8/5,4)$ under the assumption of output feedback and delayed CSIT.
To this end, we will show that in $5$ channel uses, Transmitter $1$ can send $8$ symbols to Receiver $1$ and Transmitter $2$ can send
$20$ symbols to Receiver $2$.

For all $5$ channel uses, Transmitter $2$ sends fresh information symbols, i.e., it sends
\begin{align}
X_{2}(1)=[v_{1} \ldots v_{4}]^{T},\ldots,
X_{2}(5)=[v_{17} \ldots v_{20}]^{T}.
\end{align}
In the first channel use, Transmitter $1$ sends $8$ fresh information symbols, i.e.,
\begin{align}
X_{1}(1)&=[u_{1} \hspace{0.1in} u_{2}\ldots u_{8}]^{T}.
\end{align}
Let us denote $\mathbf{u}=(u_{1},u_{2},\ldots,u_{8})$. The outputs at Receivers $1$ and $2$ at $i=1$ (ignoring noise) are given as
\begin{align}
Y_{1}(1)&=\left[
\begin{array}{c}
  A_{1}(\mathbf{u})+B_{1}(v_{1},v_{2},v_{3},v_{4}) \\
  A_{2}(\mathbf{u})+B_{2}(v_{1},v_{2},v_{3},v_{4}) \\
  A_{3}(\mathbf{u})+B_{3}(v_{1},v_{2},v_{3},v_{4}) \\
  A_{4}(\mathbf{u})+B_{4}(v_{1},v_{2},v_{3},v_{4}) \\
  A_{5}(\mathbf{u})+B_{5}(v_{1},v_{2},v_{3},v_{4}) \\
  A_{6}(\mathbf{u})+B_{6}(v_{1},v_{2},v_{3},v_{4})
\end{array}\right],\nonumber
\end{align}
\begin{align}
Y_{2}(1)&=\left[
\begin{array}{c}
  P_{1}(\mathbf{u})+Q_{1}(v_{1},v_{2},v_{3},v_{4}) \\
  P_{2}(\mathbf{u})+Q_{2}(v_{1},v_{2},v_{3},v_{4}) \\
  P_{3}(\mathbf{u})+Q_{3}(v_{1},v_{2},v_{3},v_{4})\\
  P_{4}(\mathbf{u})+Q_{4}(v_{1},v_{2},v_{3},v_{4})\\
  P_{5}(\mathbf{u})+Q_{5}(v_{1},v_{2},v_{3},v_{4})
\end{array}\right].\nonumber
\end{align}
Upon receiving feedback $Y_{1}(1)$, and channel state information $H(1)$, having access to $\mathbf{u}$, Transmitter $1$
can reconstruct $(P_{1}(\mathbf{u}),\ldots,P_{4}(\mathbf{u}))$ and $(Q_{1}(v_{1},v_{2},v_{3},v_{4}),\ldots,Q_{4}(v_{1},v_{2},v_{3},v_{4}))$.
In the subsequent channel uses, $2\leq i \leq 5$ Transmitter $1$ sends
\begin{align}
X_{1}(i)&=[P_{i-1}(\mathbf{u}) \hspace{0.1in} Q_{i-1}(v_{1},v_{2},v_{3},v_{4}) \hspace{0.1in} \phi \hspace{0.1in}\phi \hspace{0.1in}\phi \hspace{0.1in}\phi\hspace{0.1in}\phi \hspace{0.1in}\phi]^{T},\nonumber
\end{align}
where $\phi$ denotes a constant symbol known to all terminals.
It is straightforward to verify that Receiver $2$ has $25$ linearly independent equations in $25$ variables,
$(v_{1},\ldots,v_{20})$ and $(P_{1}(\mathbf{u}),\ldots,P_{5}(\mathbf{u}))$. Hence, it can decode all $20$ information symbols
$(v_{1},\ldots,v_{20})$.
\begin{figure}[t]
\centering \includegraphics[width=9.1cm]{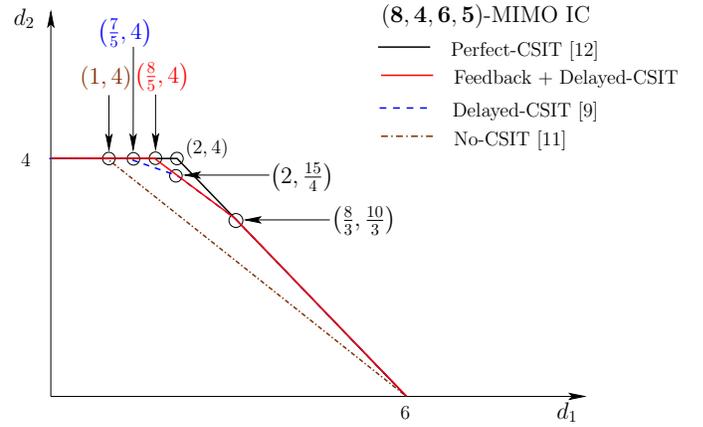}
  \caption{$\DoF$ region for the $(8,4,6,5)$-MIMO-IC with various assumptions.}\label{Figexample2}
\end{figure}

On the other hand, using $Y_{2}(i)$, Receiver $1$ can decode $P_{i}(\mathbf{u})$, and $Q_{i}(v_{1},v_{2},v_{3},v_{4})$, where $2\leq i\leq 5$.
Therefore, from $\{Y_{2}(i)\}_{i=2}^{5}$, it has $(P_{2}(\mathbf{u}),\ldots,P_{5}(\mathbf{u}))$ and $(v_{1},v_{2},v_{3},v_{4})$. Using
$(v_{1},v_{2},v_{3},v_{4})$, Receiver $1$ can reconstruct the interference signals $B_{1}(v_{1},\ldots,v_{4}),\ldots,B_{6}(v_{1},\ldots,v_{4})$
for the first channel use. Subsequently, it can subtract these and obtain $(A_{1}(\mathbf{u}),\ldots,A_{6}(\mathbf{u}))$. To summarize, Receiver $1$
can obtain $10$ equations $(A_{1}(\mathbf{u}),\ldots,A_{6}(\mathbf{u}),P_{1}(\mathbf{u}),\ldots,P_{4}(\mathbf{u}))$
 in $8$ variables and it can reliably decode $(u_{1},\ldots,u_{8})$.

 {\bf{Remark $3$}}: From Figure \ref{Figexample2}, we note that with perfect CSIT, the pair $(2,4)$ is achievable; in other words, in $5$ channel uses, one can send $10$ symbols to Receiver $1$ and $20$ symbols to Receiver $2$. However, with output feedback and delayed CSIT, to guarantee the decodability of $20$ symbols at Receiver $2$ necessitates Transmitter $1$ to repeat the interference component $(P_{1},\ldots,P_{4})$ and a part of the signal component $(Q_{1},\ldots,Q_{4})$. This coding scheme fills up all the dimensions (for this example, there are $25$) at Receiver $2$.
 However, this leaves $2$ dimensions redundant at Receiver $1$, which is the reason why feedback and delayed CSIT cannot achieve the point $(2,4)$.

\section{Conclusions}
In this paper, the $\DoF$ region of the MIMO-IC has been characterized in the presence of output feedback and delayed CSIT. It is shown that output feedback and delayed CSIT in general outperform delayed CSIT, and can sometimes be as good as perfect CSIT. Furthermore, the $\DoF$ region with local feedback and delayed CSIT is the same as the $\DoF$ region with global feedback and delayed CSIT. This implies that from the degrees of freedom perspective, local feedback yields the same performance as global feedback in the presence of delayed CSIT.   It is shown that whenever feedback and delayed CSIT strictly outperform delayed CSIT, the stronger receiver (i.e., the receiver with larger number of antennas) is able to decode both the messages. The key enabler to this  effect is the presence of feedback in addition to delayed CSIT. The converse is proved for the case in which the channels to the receivers are not necessarily identically distributed and the techniques developed herein can be useful for other related problems involving delayed CSIT.

\section{Appendix}
\subsection{Coding Scheme: arbitrary $(M_{1},M_{2},N_{1},N_{2})$}\label{Section:codingmimogeneral}
We focus on only such values of $(M_{1},M_{2},N_{1},N_{2})$ for which $\DoF^{\mathrm{d-CSIT}}\subset \DoF^{\mathrm{FB,d-CSIT}}$.
A necessary condition for this inclusion is $M_{1}>N_{1}>N_{2}>M_{2}$. For this case, the region in Theorem \ref{Theorem1} can be simplified to
\begin{align}
d_{2}&\leq M_{2}\label{B1}\\
d_{1}+d_{2}&\leq N_{1}\label{B2}\\
\frac{d_{1}}{\min(M_{1},N_{1}+N_{2})}+\frac{d_{2}}{N_{2}}&\leq 1.\label{B3}
\end{align}

We can further subdivide this scenario into two mutually exclusive cases, depending on whether the bound (\ref{B3}) is active or not:
\subsubsection{$\DoF^{\mathrm{FB,d-CSIT}}= \DoF^{\mathrm{p-CSIT}}$}\label{Section:Case-A}
In this case, the bound (\ref{B3}) is not active and hence the region is the same as that of perfect CSIT.
This condition requires $(M_{1},M_{2},N_{1},N_{2})$ to satisfy
\begin{align}
\frac{(N_{1}-M_{2})}{\min(M_{1},N_{1}+N_{2})}+\frac{M_{2}}{N_{2}}&\leq 1,
\end{align}
which is equivalent to
\begin{align}
\min(M_{1},N_{1}+N_{2})\geq N_{2}\left(\frac{N_{1}-M_{2}}{N_{2}-M_{2}}\right).
\end{align}

The $\DoF$ region with feedback and delayed CSIT is given as
\begin{align}
d_{2}&\leq M_{2}\\
d_{1}+d_{2}&\leq N_{1}.
\end{align}
The main contribution is to show the achievability of the following point:
\begin{align}
\mbox{Point }P_{0}: (N_{1}-M_{2},M_{2}).
\end{align}
The $(6,2,4,3)$-MIMO IC falls into this category.

\subsubsection{$\DoF^{\mathrm{FB,d-CSIT}}\subset \DoF^{\mathrm{p-CSIT}}$}\label{Section:Case-B}
In this case, the bound (\ref{B3}) is active and hence the region with feedback and delayed CSIT is a strict
subset of the $\DoF$ region with perfect CSIT.
This condition requires $(M_{1},M_{2},N_{1},N_{2})$ to satisfy
\begin{align}
\frac{(N_{1}-M_{2})}{\min(M_{1},N_{1}+N_{2})}+\frac{M_{2}}{N_{2}}&> 1,
\end{align}
which is equivalent to
\begin{align}
\min(M_{1},N_{1}+N_{2})< N_{2}\left(\frac{N_{1}-M_{2}}{N_{2}-M_{2}}\right).\label{caseB}
\end{align}
The $(8,4,6,5)$-MIMO IC falls into this category.

The $\DoF$ region with feedback and delayed CSIT is given as
\begin{align}
d_{2}&\leq M_{2}\\
d_{1}+d_{2}&\leq N_{1}\\
\frac{d_{1}}{\min(M_{1},N_{1}+N_{2})}+\frac{d_{2}}{N_{2}}&\leq 1.
\end{align}
The main contribution is to show the achievability of two points:
\begin{align}
\mbox{Point }P_{1}: &\Bigg[\widetilde{M}_{1}\left(\frac{N_{2}-M_{2}}{N_{2}}\right),M_{2}\Bigg],\\
\mbox{Point }P_{2}: &\Bigg[\widetilde{M}_{1}\left(\frac{N_{1}-N_{2}}{\widetilde{M}_{1}-N_{2}}\right),N_{2}\left(\frac{\widetilde{M}_{1}-N_{1}}{\widetilde{M}_{1}-N_{2}}\right)\Bigg],
\end{align}
where we have defined $\widetilde{M}_{1}=\min(M_{1},N_{1}+N_{2})$.

We have shown these two cases in Figure~\ref{FigarbitraryL}.
\begin{figure*}[t]
\centering \includegraphics[width=16.0cm]{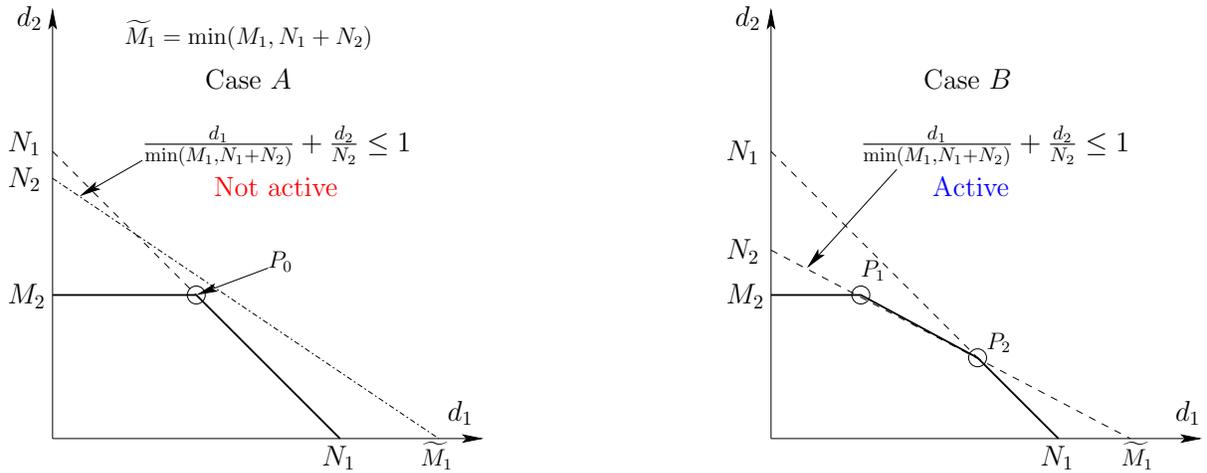}
  \caption{Two cases for the $(M_{1},M_{2},N_{1},N_{2})$-MIMO-IC.}\label{FigarbitraryL}
\end{figure*}

{\bf{Achievability for $P_{0}$ and $P_{1}$}}: Let us define $T=(N_1-M_2)(N_2-M_1)$ and 
\begin{align}
L= T\max\left(\frac{\widetilde{M}_{1}}{N_{1}-M_{2}}, \frac{N_{2}}{N_{2}-M_{2}}\right)\label{Lchoice}.
\end{align}
Note that $L$ is a positive integer. We will provide a scheme that works well for both $P_{0}$ (corresponding to Case $A$ discussed in Section~\ref{Section:Case-A}) and for Point $P_{1}$ (corresponding to Case $B$ discussed in Section~\ref{Section:Case-B}).
In particular, we will show that in $L$ channel uses, we can transmit $T\widetilde{M}_{1}$ symbols to Receiver $1$
and $LM_{2}$ symbols to Receiver $2$. Note that the technical condition distinguishing cases A and B can be
equivalently stated in terms of the value taken by the parameter $L$, i.e., $L=\frac{\widetilde{M}_{1}}{N_{1}-M_{2}}T$ for Case A, and $L=\frac{N_{2}}{N_{2}-M_{2}}T$ for Case B.

The proposed scheme includes two Phases. 
Transmitter $2$ always sends fresh information symbols in all $L$ channel uses (Phase $1$ and Phase $2$).
During the first Phase of transmission which includes time slots $i=1,\dots, T$, Transmitter $1$ sends $\widetilde{M}_{1}$ fresh information symbols in each time slot. At the end of the first Phase, Receiver $2$ has $TN_{2}$
equations, each involving an information component (a function of $TM_{2}$ information symbols) and an interference component (which are functions of $T\widetilde{M}_{1}$ symbols of Transmitter $1$).
Via feedback from Receiver $1$, Transmitter $1$ can decode the $TM_2$ information symbols of Transmitter $2$ sent over the first Phase, since $N_1\geq M_2$. Then having all the information symbols of Phase $1$ and the delayed CSIT, Transmitter $1$ can exactly recover the $TN_{2}$ interference components it caused at Receiver $2$ during Phase $1$. During Phase $2$ of the course of transmission including time slots $i=T+1,\dots, L$, Transmitter $1$ sends combinations of these interference components on its antennas, while Transmitter $2$ keeps sending $M_2$ fresh symbols in each channel use.

At the end of Phase $2$, at Receiver $2$, we have $LN_{2}$ equations in $LM_{2}+TN_{2}$ variables. Hence, for the decodability of $LM_{2}$ symbols at Receiver $2$, $L$ must satisfy
$LN_{2}\geq LM_{2}+TN_{2}$, i.e.,
\begin{align}
L\geq \frac{N_{2}}{N_{2}-M_{2}}T.
\end{align}
Furthermore, at Receiver $1$, we have $LN_{1}$ equations in $T\widetilde{M_{1}}+ LM_{2}$ variables, and hence for the decodability of $\widetilde{M}_{1}$ symbols at Receiver $1$, $L$ must satisfy $LN_{1}\geq T\widetilde{M}_{1}+LM_{2}$, i.e.,
\begin{align}
L\geq \frac{\widetilde{M}_{1}}{N_{1}-M_{2}}T.
\end{align}
Note that we choose this exact value of $L$ in (\ref{Lchoice}) to ensure the decoding requirements at both the decoders.
Consequently, we have shown the achievability of points $P_{0}$ and $P_{1}$.

{\bf{Achievability for $P_{2}$}}: Let us define
\begin{align}
L&= \widetilde{M}_{1}-N_{2}. \label{LchoiceB}
\end{align}
We will show that in $L$ channel uses, we can transmit $\widetilde{M}_{1}(N_{1}-N_{2})$ symbols to Receiver $1$
and $N_{2}(\widetilde{M}_{1}-N_{1})$ symbols to Receiver $2$.
Before proceeding, we verify the feasibility of such a scheme. Note that Transmitter $2$ has $LM_{2}$ total number of antennas (over $L$ channel uses)
 to send fresh information. Hence, for such a scheme to work, this number should exceed the total number of information symbols to be sent to Receiver $2$, i.e., we must have
\begin{align}
LM_{2}\geq N_{2}(\widetilde{M}_{1}-N_{1}),
\end{align}
which is equivalent to
\begin{align}
\widetilde{M}_{1}\geq N_{2}\left(\frac{N_{1}-M_{1}}{N_{2}-M_{1}}\right).
\end{align}
This condition is clearly satisfied from (\ref{caseB}) and the fact that $M_{1}>M_{2}$.

We now propose the coding scheme for point $P_{2}$: Transmitter $2$ sends $N_{2}(\widetilde{M}_{1}-N_{1})$ information symbols in
$L$ channel uses. Transmitter $1$ sends fresh information on $\widetilde{M}_{1}$ antennas
in the first $(N_{1}-N_{2})$ channel uses (note that $L> (N_{1}-N_{2})$). From the first $(N_{1}-N_{2})$ channel uses, upon receiving feedback and
delayed CSIT, Transmitter $1$ can reconstruct the $M_{2}(N_{1}-N_{2})$ information components and $N_{2}(N_{1}-N_{2})$ interference components of Receiver $2$.
In the subsequent $L-(N_{1}-N_{2})$ channel uses,  Transmitter $1$ forwards these two components using $M_{1}$ antennas.

Decoding at Receiver $2$: at the end of transmission, Receiver $2$ has access to $LN_{2}$ linearly independent equations in $N_{2}(\widetilde{M}_{1}-N_{1})$ information symbols
and $N_{2}(N_{1}-N_{2})$ interference components. Thus, for Receiver $2$ to decode the information symbols, we must have
\begin{align}
LN_{2}&\geq N_{2}(\widetilde{M}_{1}-N_{1})+N_{2}(N_{1}-N_{2})\\
&= N_{2}(\widetilde{M}_{1}-N_{2}).\label{decoder1B}
\end{align}

Decoding at Receiver $1$: Receiver $1$ has $LN_{1}$ equations in $\widetilde{M}_{1}(N_{1}-N_{2})$ information symbols
and $N_{2}(\widetilde{M}_{1}-N_{1})$ interference symbols. Therefore, for decoding at Receiver $1$ to succeed, we must have
\begin{align}
LN_{1}&\geq \widetilde{M}_{1}(N_{1}-N_{2})+N_{2}(\widetilde{M}_{1}-N_{1})\\
&= N_{1}(\widetilde{M}_{1}-N_{2})\label{decoder2B}.
\end{align}

Indeed we have chosen $L$ in (\ref{LchoiceB}) to satisfy both (\ref{decoder1B}) and (\ref{decoder2B}) with equality.
Thus, we have shown the achievability of the point $P_{2}$ with feedback and delayed CSIT.

 {\bf{Remark $4$}}:  As mentioned before in both Cases A and B, the $\DoF$ region of the IC with output feedback and delayed CSIT is strictly larger than that for the same channel with only delayed CSIT. This is due to the extra upper bound \label{eq:dcsi-1} which further shrinks the $\DoF$ region with only delayed CSIT when \eqref{eq:dcsi-cond-1} holds. Figure~\ref{fig:FBDCSI_vs_DCSI} illustrates the difference between the two $\DoF$ regions for both Cases A and B. It is worth mentioning that the main benefit we get from the presence of output feedback beyond delayed CSIT is that points $P_0$ (in Case A) and $P_1$ (in Case B) are achievable in the new model, while they are outside of the $\DoF$ region when there is only delayed CSIT in the system.  
\begin{figure*}[t]
\centering \includegraphics[width=16.0cm]{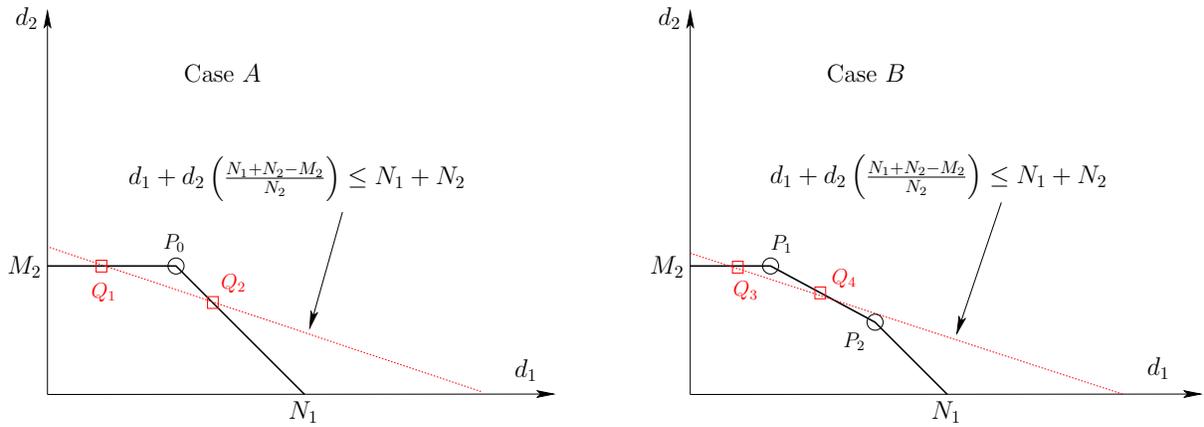}
  \caption{Comparison of $\DoF^{\mathrm{FB,d-CSIT}}$ and $\DoF^{\mathrm{d-CSIT}}$ for Cases A and B.}\label{fig:FBDCSI_vs_DCSI}
\end{figure*}

A closer look at the coding scheme presented for points $P_0$ and $P_1$ reveals additional insights towards understanding the role of output feedback and its benefit. First note that in both corner points Transmitter $2$ operates at its full rate, and keeps sending fresh information during the entire course of transmission. Therefore, Transmitter $1$ has to play two important roles simultaneously in the second Phase of transmission, i.e., during $i=T+1,\dots, L$: 
\begin{enumerate}
\item The received signal at Receiver $2$ during Phase 1 is interfered with $TN_2$ \emph{interference symbols}, each of which is an equations in terms of the $T\widetilde{M}_1$ information symbols intended for Receiver $1$. During Phase $2$, Transmitter $1$ keeps sending extra equations in terms of the same interference symbols, so that Receiver $2$ can decode  these interference symbols, and remove them from its received signal in order to decode its desired information symbols. 
\item Since $N_1>M_2$, once Receiver $1$ decodes its intended information symbols, it is able to recover the $TM_2$ information symbols intended for Receiver $2$, during Phase $1$. This implies that it not only has to recover its desired $T\widetilde{M}_1$ symbols, but also the $TM_2$ interfering symbols sent by Transmitter $1$. Hence, it requires (at least) a total of $T(\widetilde{M}_1+M_2)$ equations, while it has only received $TN_1$ of them during the first Phase. Recalling that Transmitter $2$ has to always send fresh symbols, it is Transmitter $1$ that is supposed to provide Receiver $1$ with the remaining $T(\widetilde{M}_1 + M_2 -N_2)$ equations in terms of the symbols of Phase $1$ of the course of transmission. 
\end{enumerate}
Now, note that role 1) can be accomplished by the use of only delayed CSIT, since reconstruction of the interference symbols at Transmitter $1$ requires only the channel states of Phase $1$. However, this gives only $TN_2$ equations. Since 
\[
T(\widetilde{M}_1 + M_2 -N_2) > TN_2,
\]
Transmitter $1$ requires a strictly positive number of extra equations. However, it cannot produce these extra equations from its own symbols, because it causes further interference at Receiver $2$. This is where the output feedback  plays a key role to provide Transmitter $1$ with extra equations in terms of the information symbols of Transmitter $2$ (sent during Phase $1$). By sending these equations, Transmitter $1$ can accomplish its roles simultaneously, and the $\DoF$ points $P_0$ and $P_1$ can be achieved. 

{\bf{Remark $5$}}:  It is worth mentioning that in the coding scheme proposed for points $P_0$, $P_1$, and $P_2$, always the stronger receiver (with $N_1$ antennas) is not only able to decode its own message, but also it can decode all the information symbols intended for the weaker receiver (with $N_2$ antennas). From this one can conclude that whenever $\DoF^{\mathrm{FB,d-CSIT}}$ is strictly larger that $\DoF^{\mathrm{d-CSIT}}$, the stronger receiver can decode both messages.

\subsection{Converse for MIMO IC}\label{Section:Converse}
We focus on proving (\ref{L1}) and (\ref{L2}) under assumptions A1-A3. As these bounds are symmetric, it suffices to prove (\ref{L1}). 
Before proceeding, we take a digression and prove a result for a class of cognitive interference channels with global feedback. 

\subsubsection{ Capacity of a Class of Cognitive ICs with Global Feedback}\label{Section:CIC-Feedback}
Consider the following cognitive IC, with two independent messages $W_{1}$ and $W_{2}$ intended to be decoded at Receivers $1$ and 
$2$ respectively. The channel outputs are governed by $p(y_{1},y_{2}|x_{1},x_{2})$. The message $W_2$ is available at Transmitter $2$
and both the messages $(W_{1},W_{2})$ are available at Transmitter $1$, i.e., Transmitter $1$ is cognitive. 
Furthermore, the cognitive IC is physically degraded, i.e., for any $p(x_{1},x_{2})$, the channel satisfies
\begin{align}
p(y_{1},y_{2}|x_{1},x_{2})&= p(y_{1}|x_{1},x_{2})p(y_{2}|y_{1}). \label{PDC}
\end{align}
From \cite{Wu-Viswanath}, the following region, denoted by $\mathcal{R}_{ach}$  is achievable for a general cognitive IC without feedback:
\begin{align}
R_{1}&\leq I(X_{1};Y_{1}|U,X_{2})\\
R_{2}&\leq I(U,X_{2};Y_{2}),
\end{align} 
over all probability distributions $p(x_1, x_2, u, y_1, y_2)$ that factor as
\begin{align}
p(u,x_{2})p(x_{1}|u, x_{2})p(y_{1},y_{2}|x_{1},x_{2}).
\end{align}
We now show that for a physically degraded cognitive IC that satisfies (\ref{PDC}),  the capacity region with global feedback is
given by $\mathcal{R}_{ach}$. This would imply that the capacity region without feedback is equal to the capacity region with global feedback from both receivers to both transmitters. 

We have the following sequence of bounds for $R_{1}$, the rate of message $W_{1}$:
\begin{align}
nR_{1}&= H(W_{1})\nonumber\\
&= H(W_{1}|W_{2})\nonumber\\
&\leq I(W_{1}; Y^{n}_{1}, Y^{n}_{2}|W_{2}) + n\epsilon_{1,n}\nonumber\\
&= \sum_{i=1}^{n}  I(W_{1}; Y_{1i}, Y_{2i}|W_{2}, Y^{i-1}_{1}, Y^{i-1}_{2}) + n\epsilon_{1,n}\nonumber\\
&= \sum_{i=1}^{n}  I(W_{1}; Y_{1i}, Y_{2i}|W_{2}, Y^{i-1}_{1}, Y^{i-1}_{2},X_{2i})   + n\epsilon_{1,n}  \label{E1}\\
&= \sum_{i=1}^{n}  I(W_{1}; Y_{1i}|W_{2}, Y^{i-1}_{1}, Y^{i-1}_{2},X_{2i})   \nonumber\\&\qquad+  \sum_{i=1}^{n}  I(W_{1}; Y_{2i}| Y_{1i}, W_{2}, Y^{i-1}_{1}, Y^{i-1}_{2},X_{2i})  + n\epsilon_{1,n} \nonumber\\
&= \sum_{i=1}^{n}  I(W_{1}; Y_{1i}|W_{2}, Y^{i-1}_{1}, Y^{i-1}_{2},X_{2i})   +   n\epsilon_{1,n} \label{E2}\\
&= \sum_{i=1}^{n}  I(W_{1},X_{1i}; Y_{1i}|W_{2}, Y^{i-1}_{1}, Y^{i-1}_{2},X_{2i})   +   n\epsilon_{1,n} \label{E3}\\
&= \sum_{i=1}^{n}  I(X_{1i}; Y_{1i}|W_{2}, Y^{i-1}_{1}, Y^{i-1}_{2},X_{2i})   +   n\epsilon_{1,n} \label{E4}\\
&= \sum_{i=1}^{n}  I(X_{1i}; Y_{1i}|U_{i},X_{2i})   +   n\epsilon_{1,n} \label{E5}\\
&= nI(X_{1}; Y_{1}|U,X_{2})   +   n\epsilon_{1,n} \label{E6}
\end{align}
where,
\begin{itemize}
\item (\ref{E1}) follows from the fact that $X_{2i}$ is a function of $(W_{2}, Y^{i-1}_{1}, Y^{i-1}_{2})$;
\item (\ref{E2}) follows from the physically degradedness  assumption in (\ref{PDC});
\item (\ref{E3}) follows from the fact that $X_{1i}$ is a function of $(W_{1},W_{2}, Y^{i-1}_{1}, Y^{i-1}_{2})$;
\item (\ref{E4}) follows from the memoryless property of the channel, i.e., 
\begin{align}
(W_{1},W_{2}, Y^{i-1}_{1}, Y^{i-1}_{2},) \rightarrow (X_{1i},X_{2i}) \rightarrow Y_{1i};
\end{align}
\item (\ref{E5}) follows by defining $U_{i}\triangleq (W_{2}, Y^{i-1}_{1}, Y^{i-1}_{2})$, for $i=1,\ldots,n$.
\end{itemize}
For Receiver $2$, we have the following sequence of bounds:
\begin{align}
nR_{2}&= H(W_{2})\nonumber\\
&\leq I(W_{2};Y^{n}_{2}) + n\epsilon_{2,n}\nonumber\\
&= \sum_{i=1}^{n} I(W_{2}; Y_{2i}|Y^{i-1}_{2}) + n\epsilon_{2,n}\nonumber\\
&\leq  \sum_{i=1}^{n} I(W_{2},Y^{i-1}_{1}, Y^{i-1}_{2}; Y_{2i}) + n\epsilon_{2,n}\nonumber\\
&=  \sum_{i=1}^{n} I(W_{2},Y^{i-1}_{1}, Y^{i-1}_{2}, X_{2i}; Y_{2i}) + n\epsilon_{2,n}\label{E7}\\
&=  \sum_{i=1}^{n} I(U_{i}, X_{2i}; Y_{2i}) + n\epsilon_{2,n}\label{E8}\\
&=  nI(U, X_{2}; Y_{2}) + n\epsilon_{2,n}\label{E9}
\end{align}
where
\begin{itemize}
\item (\ref{E7}) follows from the fact that $X_{2i}$ is a function of $(W_{2}, Y^{i-1}_{1}, Y^{i-1}_{2})$; and
\item (\ref{E8}) follows from the definition of $U_{i}\triangleq (W_{2}, Y^{i-1}_{1}, Y^{i-1}_{2})$, for $i=1,\ldots,n$.
\end{itemize}
From (\ref{E6}) and (\ref{E9}), we have the bounds
\begin{align}
R_{1}&\leq I(X_{1};Y_{1}|U,X_{2})\\
R_{2}&\leq I(U,X_{2};Y_{2}),
\end{align}
and it is straightforward to check that the distribution of the variables $p(x_1, x_2, u, y_1, y_2)$  satisfies
\begin{align}
p(u,x_{2})p(x_{1}|u, x_{2})p(y_{1}|x_{1},x_{2})p(y_{2}|y_{1}).
\end{align}
Therefore, we have shown that global output feedback does not increase the capacity region of the physically degraded cognitive interference channel\footnote{We note here that if for the cognitive IC, the physical degradedness order is switched, i.e., when $Y_{1}$ is a degraded version of $Y_{2}$, then the capacity regions with and without feedback are not known in general.}. 
We note here that this result can be regarded as cognitive interference channel counterpart of the corresponding result for the physically degraded 
broadcast channel, for which is it known \cite{ElGamalFB} that feedback does not increase the capacity region. 

\subsubsection{Proof of (\ref{L1})}
\label{sec:L1}
We now return to the $(M_{1},M_{2}, N_{1}, N_{2})$ MIMO IC. We will prove an outer bound with global feedback and delayed CSIT.
To this end, we provide the following enhancement of the original MIMO IC (denote it as O-IC):
\begin{itemize}
\item Provide the message $W_{2}$ to Transmitter $1$.
\item Provide the channel output $Y_{2}$ to Receiver $1$.
\end{itemize}
Thus, we now have an enhanced MIMO IC, (denote it as E-IC) summarized as follows:
\begin{itemize}
\item Transmitter $1$ has $(W_{1},W_{2})$, with global feedback and delayed CSIT.
\item Transmitter $2$ has $W_{2}$, with global feedback and delayed CSIT.
\item Receiver $1$ has $(Y^{n}_{1},Y^{n}_{2},\mathbf{H}^{n})$.
\item Receiver $2$ has $(Y^{n}_{2},\mathbf{H}^{n})$.
\end{itemize}
Clearly, this enhanced MIMO IC (E-IC) falls into the class of physically degraded cognitive ICs, for which we have shown in the previous section (\ref{Section:CIC-Feedback}) that
{\emph{global feedback}} does not increase the capacity region. Thus, we can remove the global feedback and delayed CSIT assumption from the enhanced MIMO IC
without changing its capacity region. 

We thus consider the following MIMO cognitive IC (denote it as E-IC$^{*}$) in which there is no feedback and no CSIT:
\begin{itemize}
\item Transmitter $1$ has $(W_{1},W_{2})$, with no feedback and no CSIT.
\item Transmitter $2$ has $W_{2}$, with no feedback and no CSIT.
\item Receiver $1$ has $(Y^{n}_{1},Y^{n}_{2},\mathbf{H}^{n})$.
\item Receiver $2$ has $(Y^{n}_{2},\mathbf{H}^{n})$.
\end{itemize}
In particular, the following encoding functions are permissible for  E-IC$^{*}$:
\begin{align}
X_{1}(i)= f_{1i}(W_{1},W_{2}),\quad  X_{2}(i)= f_{2i}(W_{2}).\label{EncodingEIC}
\end{align}
Let us denote
\begin{align}
&M_{u}=\min(M_{1},N_{1}+N_{2}), \quad M_{l}=\min(M_{1},N_{2}), \nonumber\\
&\hspace{2.5cm}M^{*}= M_{u}-M_{l}.
\end{align}
We next create an artificial  channel output (of size $M^{*}\times 1$) as follows:
\begin{align}
\widetilde{Y}(i)= \mathbf{\widetilde{H}}(i)X_{1}(i)+  \widetilde{Z}(i),\label{artificial}
\end{align}
where 
\begin{itemize}
\item The channel matrix $\mathbf{\widetilde{H}}(i)$ is of size $M^{*}\times M_{1}$.
\item Each element of $\mathbf{\widetilde{H}}(i)$ is distributed i.i.d. from the distribution $\lambda_{21}$.
\item The realizations of $\mathbf{\widetilde{H}}(i)$ vary in an i.i.d. manner over time. 
\item The elements of $\widetilde{Z}(i)$ are generated i.i.d. from $\mathcal{CN}(0,1)$, and vary in an i.i.d. manner across time.
\end{itemize}
We collectively denote the channel state information of the original MIMO IC and the channel of the $M^{*}\times 1$ artificial channel output as follows:
\begin{align}
\Omega&= \{\mathbf{H}_{11}(i),\mathbf{H}_{12}(i), \mathbf{H}_{21}(i), \mathbf{H}_{22}(i), \mathbf{\widetilde{H}}(i)\}_{i=1}^{n}.
\end{align}
We have the following sequence of inequalities for Receiver $2$:
\begin{align}
nR_{2}&= H(W_{2})\nonumber\\
&= H(W_{2}|\Omega)\nonumber\\
&\leq I(W_{2};Y^{n}_{2}|\Omega) + n\epsilon_{2,n}\nonumber\\
&= \sum_{i=1}^{n} I(W_{2}; Y_{2}(i)| Y^{i-1}_{2}, \Omega) + n\epsilon_{2,n}\nonumber\\
&\leq \sum_{i=1}^{n} I(W_{2}, Y^{i-1}_{2}; Y_{2}(i)|\Omega) + n\epsilon_{2,n}\nonumber\\
&\leq \sum_{i=1}^{n} I(W_{2}, Y^{i-1}_{1},Y^{i-1}_{2}, \widetilde{Y}^{i-1}; Y_{2}(i)|\Omega) + n\epsilon_{2,n}\nonumber\\
&= \sum_{i=1}^{n}h(Y_{2}(i)|\Omega) \nonumber\\&\quad- \sum_{i=1}^{n} h(Y_{2}(i)|\Omega, W_{2}, Y^{i-1}_{1},Y^{i-1}_{2}, \widetilde{Y}^{i-1}) + n\epsilon_{2,n}\nonumber
\end{align}
\begin{align}
&\leq n\min(N_{2},M_{1}+M_{2})\log(P) \nonumber\\&\quad- \sum_{i=1}^{n} h(Y_{2}(i)|\Omega, W_{2}, Y^{i-1}_{1},Y^{i-1}_{2}, \widetilde{Y}^{i-1}) + n\epsilon_{2,n}\nonumber\\
&= n\min(N_{2},M_{1}+M_{2})\log(P) \nonumber\\&\quad- \sum_{i=1}^{n} h(Y_{2}(i)|\Omega, W_{2}, X_{2}(i), Y^{i-1}_{1},Y^{i-1}_{2}, \widetilde{Y}^{i-1}) + n\epsilon_{2,n}\label{EQ1},
\end{align}
where in (\ref{EQ1}) we used the fact from (\ref{EncodingEIC}) that $X_{2}(i)$ is a function of $W_{2}$.

We next have the following sequence of inequalities for Receiver $1$:
\begin{align}
&nR_{1}\nonumber\\
&= H(W_{1})\nonumber\\
&= H(W_{1}|\Omega,W_{2})\nonumber\\
&\leq I(W_{1}; Y^{n}_{1}, Y^{n}_{2}, \widetilde{Y}^{n}|\Omega,W_{2}) + n\epsilon_{1,n}\nonumber\\
&= \sum_{i=1}^{n} I(W_{1}; Y_{1}(i), Y_{2}(i), \widetilde{Y}(i)| \Omega,W_{2}, Y^{i-1}_{1},Y^{i-1}_{2}, \widetilde{Y}^{i-1}) \nonumber\\&\quad+ n\epsilon_{1,n}\nonumber\\
&= \sum_{i=1}^{n}h(Y_{1}(i), Y_{2}(i), \widetilde{Y}(i)| \Omega,W_{2}, Y^{i-1}_{1},Y^{i-1}_{2}, \widetilde{Y}^{i-1})  \nonumber\\
&\quad - \sum_{i=1}^{n}\underbrace{h(Y_{1}(i), Y_{2}(i), \widetilde{Y}(i)| \Omega,W_{1}, W_{2}, Y^{i-1}_{1},Y^{i-1}_{2}, \widetilde{Y}^{i-1})}_{\geq o(\log(P))} \nonumber\\&\quad+ n\epsilon_{1,n}\nonumber\\
&\leq \sum_{i=1}^{n}h(Y_{1}(i), Y_{2}(i), \widetilde{Y}(i) | \Omega,W_{2}, Y^{i-1}_{1},Y^{i-1}_{2}, \widetilde{Y}^{i-1})  + n\epsilon_{1,n} \label{EQ1b}\\
&= \sum_{i=1}^{n} h(Y_{2}(i),\widetilde{Y}(i)|\Omega, W_{2}, X_{2}(i), Y^{i-1}_{1},Y^{i-1}_{2}, \widetilde{Y}^{i-1})  \nonumber\\
&\quad + \sum_{i=1}^{n} \underbrace{h(Y_{1}(i)|\Omega, W_{2}, X_{2}(i), Y_{2}(i), \widetilde{Y}(i), Y^{i-1}_{1},Y^{i-1}_{2}, \widetilde{Y}^{i-1})}_{\leq o(\log(P))} \nonumber\\&\quad+ n\epsilon_{1,n}\nonumber\\
&\leq \sum_{i=1}^{n} h(Y_{2}(i),\widetilde{Y}(i) |\Omega, W_{2}, X_{2}(i), Y^{i-1}_{1},Y^{i-1}_{2}, \widetilde{Y}^{i-1})   \nonumber\\&\quad + n(o(\log(P))+\epsilon_{1,n})\label{EQ2},
\end{align}
where
\begin{itemize}
\item (\ref{EQ1b}) follows from the fact that $(Y_{1}(i),Y_{2}(i),\widetilde{Y}(i))$ can be obtained within noise distortion from $(\Omega, W_{1}, W_{2})$; and
\item (\ref{EQ2}) follows from the fact that $Y_{1}(i)$ can be obtained within  noise distortion \break from $(\Omega, W_{2}, X_{2}(i), Y_{2}(i), \widetilde{Y}(i))$ via channel inversion. 
\end{itemize}

The next key step is to relate the quantities in equations (\ref{EQ1}) and (\ref{EQ2}). For simplicity, define a variable
\begin{align}
\Sigma(i)&\triangleq \left(\Omega, W_{2}, X_{2}(i), Y^{i-1}_{1},Y^{i-1}_{2}, \widetilde{Y}^{i-1}\right),
\end{align}
which appears in the conditioning of both summations appearing in (\ref{EQ1}) and (\ref{EQ2}). 
Using this definition, we can compactly re-write (\ref{EQ1}) and (\ref{EQ2}) as follows:
\begin{align}
nR_{2}&\leq n\min(N_{2},M_{1}+M_{2})\log(P) \nonumber\\&\quad- \sum_{i=1}^{n} h(Y_{2}(i)|\Sigma(i)) + n\epsilon_{2,n}\label{EQA}\\
nR_{1}&\leq \sum_{i=1}^{n} h(Y_{2}(i),\widetilde{Y}(i)|\Sigma(i)) + n(o(\log(P))+\epsilon_{1,n}).\label{EQB}
\end{align}
Before proceeding, we set the following notation: 
\begin{itemize}
\item $Y_{2}(i,q)$ denotes the output at the $q$th antenna at Receiver $2$ at time $i$. 
\item $\widetilde{Y}(i,r)$ denotes the output at the $r$th antenna of the artificial channel output at time $i$.
\end{itemize}
We next describe the consequence of the statistical equivalence of the channel $\mathbf{\widetilde{H}}(i)$ used to define the artificial channel output in (\ref{artificial}).
In particular, consider the outputs of Receiver $2$ and artificial channel output.

We now claim the following statistically equivalent property (SEP): 
\begin{align}
\mbox{(SEP)}: h(Y_{2}(i, q)|\Sigma(i)) = h(\widetilde{Y}(i, r)|\Sigma(i)) \label{SEP}
\end{align}
for any $q\in \{1,\ldots, N_{2}\}$ and any $r\in \{1,\ldots, M^{*}\}$. 
Note that the key is that $\Sigma(i)$ is within the conditioning which allows us to use the statistical equivalence. In particular, 
$\Sigma(i)$ comprises $(W_{2},\Omega)$ and $X_{2}(i)$ is a function of $W_{2}$. Therefore, conditioned on $\Sigma(i)$, the contribution of $\mathbf{H}_{22}(i)X_{2}(i)$ can be subtracted 
from $Y_{2}(i)$ given $\Sigma(i)$. In particular, we have, 
\begin{align}
h(Y_{2}(i)|\Sigma(i))&= h\left(\mathbf{H}_{21}(i)X_{1}(i)+ Z_{2}(i) | \Sigma(i)\right)\\
h(\widetilde{Y}(i)|\Sigma(i))&= h\left(\mathbf{\widetilde{H}}(i)X_{1}(i)+ \widetilde{Z}(i) | \Sigma(i)\right).
\end{align}
Furthermore, all elements of $\mathbf{H}_{21}(i)$ and $\mathbf{\widetilde{H}}(i)$ are generated i.i.d. from the {\emph{same}} distribution $\lambda_{21}$.
This in turn implies that given $\Sigma(i)$, $Y_{2}(i,q)$ and $\widetilde{Y}(i,r)$ are identically distributed, and thus (\ref{SEP}) follows.

Now, consider the term appearing in the summation in (\ref{EQA}):
\begin{align}
&h(Y_{2}(i)|\Sigma(i))\nonumber\\
&= \sum_{q=1}^{N_{2}} h(Y_{2}(i, q) | \Sigma(i), Y_{2}(i, 1), \ldots,Y_{2}(i, q-1)) \nonumber\\
&= \sum_{q=1}^{M_{l}} h(Y_{2}(i, q) | \Sigma(i), Y_{2}(i, 1), \ldots,Y_{2}(i, q-1))  \nonumber\\
&\qquad + \sum_{q=M_{l}+1}^{N_{2}} \underbrace{h(Y_{2}(i, q) | \Sigma(i), Y_{2}(i, 1), \ldots,Y_{2}(i, q-1))}_{\geq o(\log(P))} \nonumber\\
&\geq \sum_{q=1}^{M_{l}} h(Y_{2}(i, q) | \Sigma(i), Y_{2}(i, 1), \ldots,Y_{2}(i, q-1))  \nonumber\\
&= \sum_{q=1}^{M_{l}} h(Y_{2}(i, \min(M_{1},N_{2})) | \Sigma(i), Y_{2}(i, 1), \ldots,Y_{2}(i, q-1))  \label{EQ3}
\end{align}
\begin{align}
&\geq \sum_{q=1}^{M_{l}} h(Y_{2}(i, M_{l}) | \Sigma(i), Y_{2}(i, 1), \ldots,Y_{2}(i, M_{l}-1))  \nonumber\\
&= M_{l} \cdot h(Y_{2}(i, M_{l}) | \Sigma(i), Y_{2}(i, 1), \ldots,Y_{2}(i, M_{l}-1)) \\
&\triangleq M_{l}\cdot \zeta(i), \label{EQ4}
\end{align}
where in (\ref{EQ3}) we made use of the statistically equivalent property (SEP), and in (\ref{EQ4}), we have defined
\begin{align}
\zeta(i)&\triangleq h(Y_{2}(i, M_{l})) | \Sigma(i), Y_{2}(i, 1), \ldots,Y_{2}(i, M_{l}-1)),
\end{align}
and note that $M_{l}=\min(M_{1},N_{2})$. 

We next consider the term appearing in the summation in (\ref{EQB}):
\begin{align}
&h(Y_{2}(i),\widetilde{Y}(i)|\Sigma(i)) \nonumber\\
&= h(Y_{2}(i)|\Sigma(i)) + h(\widetilde{Y}(i)|\Sigma(i), Y_{2}(i)) \nonumber\\
&\leq h(Y_{2}(i)|\Sigma(i)) \nonumber\\&\quad+ h(\widetilde{Y}(i)| \Sigma(i), Y_{2}(i,1),\ldots, Y_{2}(i, \min(M_{1},N_{2}))) \nonumber\\
&= h(Y_{2}(i)|\Sigma(i)) \nonumber\\&\quad+ h(\tilde{Y}(i,1),.., \tilde{Y}(i, M_{u}-M_{l}) | \Sigma(i), Y_{2}(i,1),.., Y_{2}(i, M_{l}))) \nonumber\\
&= h(Y_{2}(i)|\Sigma(i)) \nonumber\\&\quad+ \sum_{r=1}^{M_{u}-M_{l}}h(\tilde{Y}(i,r) | \Sigma(i), Y_{2}(i,1),\ldots, Y_{2}(i, M_{l}))) \nonumber\\
&\leq h(Y_{2}(i)|\Sigma(i)) \nonumber\\&\quad+ \sum_{r=1}^{M_{u}-M_{l}}h(\tilde{Y}(i,r) | \Sigma(i), Y_{2}(i,1),\ldots, Y_{2}(i, M_{l}-1))) \nonumber\\
&= h(Y_{2}(i)|\Sigma(i)) \nonumber\\&\quad+ \sum_{r=1}^{M_{u}-M_{l}}h(Y_{2}(i,M_{l})) | \Sigma(i), Y_{2}(i,1),\ldots, Y_{2}(i, M_{l}-1))) \label{Use2}\\
&= h(Y_{2}(i)|\Sigma(i)) + (M_{u}-M_{l})\zeta(i) \nonumber\\
&= h(Y_{2}(i)|\Sigma(i)) + (\min(M_{1},N_{1}+N_{2})-\min(M_{1},N_{2}))\zeta(i)\label{EQ5},
\end{align}
where (\ref{Use2}) follows from the SEP property.

From (\ref{EQ4}) and (\ref{EQ5}), we eliminate $\zeta(i)$, to obtain
\begin{align}
h(Y_{2}(i),\widetilde{Y}(i)|\Sigma(i))&\leq \frac{\min(M_{1},N_{1}+N_{2})}{\min(M_{1},N_{2})}h(Y_{2}(i)|\Sigma(i)). \label{REL}
\end{align}
From (\ref{EQA}), (\ref{EQB}), and (\ref{REL}), we obtain
\begin{align}
nR_{2}&\leq n\min(N_{2},M_{1}+M_{2})\log(P) - \sum_{i=1}^{n} h(Y_{2}(i)|\Sigma(i)) \nonumber\\&\quad+ n\epsilon_{2,n}\\
nR_{1}&\leq\frac{\min(M_{1},N_{1}+N_{2})}{\min(M_{1},N_{2})}\sum_{i=1}^{n} h(Y_{2}(i)|\Sigma(i)) \nonumber\\&\quad+ n(2o(\log(P)))+\epsilon_{1,n}),
\end{align}
which imply that 
\begin{align}
&R_{1}+ \frac{\min(M_{1},N_{1}+N_{2})}{\min(M_{1},N_{2})}R_{2}\nonumber\\
&\quad \leq \frac{\min(N_{2},M_{1}+M_{2})\min(M_{1},N_{1}+N_{2})}{\min(M_{1},N_{2})}\log(P)  \nonumber\\&\quad\quad+ (2o(\log(P))+ \epsilon_{n}).
\end{align}
Dividing by $\log(P)$ and taking the limits $n\rightarrow \infty$ and then $P\rightarrow \infty$, we have the proof for
\begin{align}
\frac{d_{1}}{\min(M_{1},N_{1}+N_{2})} + \frac{d_{2}}{\min(M_{1},N_{2})} &\leq  \frac{\min(N_{2},M_{1}+M_{2})}{\min(M_{1},N_{2})}.
\end{align}

\section*{Acknowledgement}
We are grateful to the Associate Editor Syed. A. Jafar and the anonymous reviewers for their suggestions which
led to the enhancement of the scope and the readability of the material.

%
%

%

\begin{IEEEbiography}{Ravi Tandon} (SÕ03, MÕ09) received the B.Tech degree in electrical engineering from the Indian Institute of Technology (IIT), Kanpur in 2004 and the Ph.D. degree in electrical and computer engineering from the University of Maryland, College Park in 2010. From 2010 until 2012, he was a post-doctoral research associate with Princeton University. In 2012, he joined Virginia Polytechnic Institute and State University (Virginia Tech) at Blacksburg, where he is currently a Research Assistant Professor in the Department of Electrical and Computer Engineering. His research interests are in network information theory, communication theory for wireless networks and information theoretic security.

Dr. Tandon is a recipient of the Best Paper Award at the Communication Theory symposium at the 2011 IEEE Global Telecommunications Conference. 
\end{IEEEbiography}

\begin{IEEEbiography}{Soheil Mohajer} received the B.Sc. degree in electrical engineering from the Sharif University of Technology, Tehran, Iran, in 2004, and the M.Sc. and Ph.D. degrees in communication systems both from Ecole Polytechnique F\'ed\'erale de Lausanne (EPFL), Lausanne, Switzerland, in 2005 and 2010, respectively. He then joined Princeton University, New Jersey, as a post-doctoral research associate. Dr. Mohajer has been a post-doctoral researcher at the University of California at Berkeley, since October 2011.

His research interests include multi-user information theory, statistical machine learning, and bioinformatics.
\end{IEEEbiography}

\begin{IEEEbiography}{H. Vincent Poor}
(SÕ72, MÕ77, SMÕ82, FÕ87) received the Ph.D. degree in electrical engineering and computer science from Princeton University in 1977.  From 1977 until 1990, he was on the faculty of the University of Illinois at Urbana-Champaign. Since 1990 he has been on the faculty at Princeton, where he is the Dean of Engineering and Applied Science, and the Michael Henry Strater University Professor of Electrical Engineering. Dr. Poor's research interests are in the areas of stochastic analysis, statistical signal processing and information theory, and their applications in wireless networks and related fields including social networks and smart grid. Among his publications in these areas are {\em{Classical, Semi-classical and Quantum Noise}} (Springer, 2012) and {\em{Smart Grid Communications and Networking}} (Cambridge University Press, 2012).

Dr. Poor is a member of the National Academy of Engineering and the National Academy of Sciences, a Fellow of the American Academy of Arts and Sciences, and an International Fellow of the Royal Academy of Engineering (U. K.). He is also a Fellow of the Institute of Mathematical Statistics, the Optical Society of America, and other organizations.  In 1990, he served as President of the IEEE Information Theory Society, in 2004-07 as the Editor-in-Chief of these {\em{Transactions}}, and in 2009 as General Co-chair of the IEEE International Symposium on Information Theory, held in Seoul, South Korea. He received a Guggenheim Fellowship in 2002 and the IEEE Education Medal in 2005. Recent recognition of his work includes the 2010 IET Ambrose Fleming Medal for Achievement in Communications, the 2011 IEEE Eric E. Sumner Award, the 2011 IEEE Information Theory Paper Award, and honorary doctorates from Aalborg University, the Hong Kong University of Science and Technology, and the University of Edinburgh.
\end{IEEEbiography}

\begin{IEEEbiography}{Shlomo Shamai (Shitz)} received the B.Sc., M.Sc., and Ph.D. degrees in
electrical engineering from the Technion---Israel Institute of Technology,
in 1975, 1981 and 1986 respectively.

During 1975-1985 he was with the Communications Research Labs,
in the capacity of a Senior Research Engineer. Since 1986 he is with
the Department of Electrical Engineering, Technion---Israel Institute of
Technology, where he is now a Technion Distinguished Professor,
and holds the William Fondiller Chair of Telecommunications.
His research interests encompasses a wide spectrum of topics in information
theory and statistical communications.

Dr. Shamai (Shitz) is an IEEE Fellow and a member of the Israeli Academy of
Sciences and Humanities. He is the recipient of the 2011
Claude E. Shannon Award.
He has been awarded the 1999 van der Pol Gold Medal of the Union Radio
Scientifique Internationale (URSI), and is a co-recipient of the 2000 IEEE
Donald G. Fink Prize Paper Award, the 2003, and
the 2004 joint IT/COM societies paper award, the 2007 IEEE Information
Theory Society Paper Award, the 2009 European Commission FP7, Network of
Excellence in Wireless COMmunications (NEWCOM++) Best Paper Award,
and the 2010 Thomson Reuters Award for International Excellence
in Scientific Research.  He is also the recipient of
1985 Alon Grant for distinguished young scientists and the 2000 Technion Henry
Taub Prize for Excellence in Research.
He has served as Associate Editor for the Shannon Theory of the IEEE
Transactions on Information Theory, and has also served twice on the
Board of Governors of the Information Theory Society.
He is a member of the Executive Editorial Board of the IEEE Transactions
on Information Theory
\end{IEEEbiography}


\begin{thebibliography}{10}

\bibitem{Jafar:alignmenet}
V.~R. Cadambe and S.~A. Jafar.
\newblock Interference alignment and degrees of freedom of the {K}-user
  interference channel.
\newblock {\em IEEE Trans. Inf. Theory}, 54(8):3425--3441, Aug. 2008.

\bibitem{MaddahAli:X}
M.~A. Maddah-Ali, A.~S. Motahari, and A.~K. Khandani.
\newblock Communication over {M}{I}{M}{O} {X} channels: Interference alignment,
  decomposition, and performance analysis.
\newblock {\em IEEE Trans. Inf. Theory}, 54(8):3457--3470, Aug. 2008.

\bibitem{mohajer:neutralization}
S.~Mohajer, S.~Diggavi, C.~Fragouli, and D.~N.~C. Tse.
\newblock Approximate capacity characterization for a class of {G}aussian
  relay-interference wireless networks.
\newblock {\em IEEE Trans. Inf. Theory}, 57(5):2837--2864, May 2011.

\bibitem{Jafar:neutralization}
T.~Gou, S.~A. Jafar, S.~W. Jeon, and S.~Y. Chung.
\newblock Aligned interference neutralization and the degrees of freedom of the
  2$\times$2$\times$2 interference channel.
\newblock {\em IEEE Trans. Inf. Theory}, 58(7):4381--4395, July 2012.

\bibitem{Jafar:Tutorial}
S.~A. Jafar.
\newblock Interference alignment: A new look at signal dimensions in a
  communication network.
\newblock {\em Foundations and Trends in Communications and Information
  Theory}, 7(1):1--134, 2010.

\bibitem{MaddahAli-Tse:DCSI-BC}
M.~A. Maddah-Ali and D.~Tse.
\newblock Completely stale transmitter channel state information is still very
  useful.
\newblock {\em IEEE Trans. Inf. Theory}, 58(7):4418--4431, July 2012.

\bibitem{Vaze:MIMOBC}
C.~S. Vaze and M.~K. Varanasi.
\newblock The degrees of freedom region of the two-user {M}{I}{M}{O} broadcast
  channel with delayed {C}{S}{I}.
\newblock {\em Arxiv preprint arXiv:1101.0306}, 2010.

\bibitem{Retro-IA}
H.~Maleki, S.~A. Jafar, and S.~Shamai.
\newblock Retrospective interference alignment over interference networks.
\newblock {\em IEEE Journal of Selected Topics in Signal Processing},
  6(3):228--240, June 2012.

\bibitem{Vaze:MIMOICDelayedCSIT}
C.~S. Vaze and M.~K. Varanasi.
\newblock The degrees of freedom region and interference alignment for the
  {M}{I}{M}{O} interference channel with delayed {C}{S}{I}{T}.
\newblock {\em IEEE Trans. Inf. Theory}, 58(7):4396--4417, July 2012.

\bibitem{Vaze:NoCSIT-A}
C.~S. Vaze and M.~K. Varanasi.
\newblock The degree-of-freedom regions of {M}{I}{M}{O} broadcast,
  interference, and cognitive radio channels with no {C}{S}{I}{T}.
\newblock {\em IEEE Trans. Inf. Theory}, 58(8):5354--5374, Sept. 2012.

\bibitem{Vaze:NoCSIT-B}
C.~S. Vaze and M.~K. Varanasi.
\newblock A new outer-bound via interference localization and the degrees of
  freedom regions of {M}{I}{M}{O} interference networks with no {C}{S}{I}{T}.
\newblock {\em IEEE Trans. Inf. Theory, \emph{to appear}}.

\bibitem{JafarFakhereddin:MIMOIC}
S.~A. Jafar and M.~Fakhereddin.
\newblock Degrees of freedom for the {M}{I}{M}{O} interference channel.
\newblock {\em IEEE Trans. Inf. Theory}, 53(7):2637--2642, July 2007.

\bibitem{zero:56}
C.~E. Shannon.
\newblock The zero error capacity of a noisy channel.
\newblock {\em IRE Trans. Inf. Theory}, 2(3):8--19, 1956.

\bibitem{GW:1975}
N.~Gaarder and J.~Wolf.
\newblock The capacity region of a multiple-access discrete memoryless channel
  can increase with feedback (corresp.).
\newblock {\em IEEE Trans. Inf. Theory}, 21(1):100--102, Jan. 1975.

\bibitem{Ozarow:BCFB}
L.~Ozarow and S.~Leung-Yan-Cheong.
\newblock An achievable region and outer bound for the {G}aussian broadcast
  channel with feedback (corresp.).
\newblock {\em IEEE Trans. Inf. Theory}, 30(4):667--671, July 1984.

\bibitem{SuhTseIT}
C.~Suh and D.~Tse.
\newblock Feedback capacity of the {G}aussian interference channel to within
  two bits.
\newblock {\em IEEE Trans. Inf. Theory}, 57(5):2667--2685, May 2011.

\bibitem{prabhakaran2011interference}
V.M. Prabhakaran and P.~Viswanath.
\newblock Interference channels with source cooperation.
\newblock {\em IEEE Trans. Inf. Theory}, 57(1):156--186, Jan. 2011.

\bibitem{CadambeJafar2009:cooperation}
V.~R. Cadambe and S.A. Jafar.
\newblock Degrees of freedom of wireless networks with relays, feedback,
  cooperation and full duplex operation.
\newblock {\em IEEE Trans. Inf. Theory}, 55(5):2334--2344, May 2009.

\bibitem{HuangJafar2009}
C.~Huang and S.A. Jafar.
\newblock Degrees of freedom of the {M}{I}{M}{O} interference channel with
  cooperation and cognition.
\newblock {\em IEEE Trans. Inf. Theory}, 55(9):4211--4220, Sept. 2009.

\bibitem{Vaze:ShannonFB}
C.~S. Vaze and M.~K. Varanasi.
\newblock The degrees of freedom region of the {M}{I}{M}{O} interference
  channel with {S}hannon feedback.
\newblock {\em Arxiv preprint arXiv:1109.5779}, 2012.

\bibitem{GouJafar:MixedCSIT}
T.~Gou and S.~A. Jafar.
\newblock Optimal use of current and outdated channel state information --
  degrees of freedom of the {M}{I}{S}{O} {B}{C} with mixed {C}{S}{I}{T}.
\newblock {\em IEEE Commun. Lett.}, 16(7):1084--1087, July 2012.

\bibitem{ICC2012Tandon}
R.~Tandon, S.~Mohajer, H.~V. Poor, and S.~Shamai.
\newblock Feedback and delayed {C}{S}{I} can be as good as perfect {C}{S}{I}.
\newblock In {\em Proc. IEEE International Conference on Communications (ICC)},
  Ottawa, Canada, 2012.

\bibitem{Wu-Viswanath}
W.~Wu, S.~Vishwanath, and A.~Arapostathis.
\newblock Capacity of a class of cognitive radio channels: Interference
  channels with degraded message sets.
\newblock {\em IEEE Trans. Inf. Theory}, 53(11):4391--4399, Nov. 2007.

\bibitem{ElGamalFB}
A.~El Gamal.
\newblock The feedback capacity of degraded broadcast channels.
\newblock {\em IEEE Trans. Inf. Theory}, 24(3):379--381, May 1978.

\end{thebibliography}
\end{document}